\documentclass[prd,twocolumn,floatfix,nofootinbib,superscriptaddress,showpacs]{revtex4}
\pdfoutput=1
\usepackage{graphicx}
\usepackage{epsfig}
\usepackage{bm}
\usepackage{amssymb}
\usepackage{float}
\usepackage{amsmath}
\usepackage{dcolumn}
\usepackage{cancel}
\usepackage[colorlinks]{hyperref}
\usepackage[usenames,dvipsnames]{color}
\hypersetup{
     breaklinks=true,
    pdfstartview={FitH},    
    colorlinks=true,       
    linkcolor=blue,          
    citecolor=red,        
    filecolor=magenta,      
    urlcolor=blue,           
    anchorcolor=green,      
    linktocpage=true
}

\def\b{ \beta}
\def\g{ \gamma}
\def\d{ \delta}
\def\m{ \mu}
\def\n{ \nu}

\def\r{ \rho}
\def\s{ \sigma}

\usepackage{amsmath,latexsym}
\usepackage{placeins}
\usepackage[toc,page]{appendix}

\newcommand{\mathsym}[1]{{}}
\newcommand{\unicode}[1]{{}}
\newcommand{\be}{\begin{equation}}
\newcommand{\ee}{\end{equation}}
\newcommand{\bea}{\begin{eqnarray}}
\newcommand{\eea}{\end{eqnarray}}
\newcommand{\beaa}{\begin{eqnarray*}}
\newcommand{\eeaa}{\end{eqnarray*}}

\begin{document}

\preprint[\leftline{KCL-PH-TH/2021-{\bf 48}}

\vspace{0.2cm}

\title{Big Bang Nucleosynthesis constraints on higher-order modified gravities}

\author{Petros Asimakis}
\affiliation{Department of Physics, School of Applied Mathematical and Physical Sciences, National Technical University of Athens, 9 Iroon Polytechniou Str., Zografou Campus GR 157 80, Athens, Greece}

\author{Spyros Basilakos}
\affiliation{National Observatory of Athens, Lofos Nymfon, 11852 Athens, Greece}
\affiliation{Academy of Athens, Research Center for Astronomy and Applied 
Mathematics,\\
Soranou Efesiou 4, 11527, Athens, Greece}
\affiliation{
School of Sciences, European University Cyprus, Diogenes Street, Engomi 1516 
Nicosia}

\author{Nick E. Mavromatos}

\affiliation{Department of Physics, School of Applied Mathematical and Physical Sciences, National Technical University of Athens, 9 Iroon Polytechniou Str., Zografou Campus GR 157 80, Athens, Greece}

\affiliation{Theoretical Particle Physics and Cosmology Group,
Physics Department,
King’s College London,
Strand, London WC2R 2LS}

\author{Emmanuel N. Saridakis}
\affiliation{National Observatory of Athens, Lofos Nymfon, 11852 Athens,
Greece}
\affiliation{CAS Key Laboratory for Researches in Galaxies and Cosmology,
Department of Astronomy, University of Science and Technology of China, Hefei,
Anhui 230026, P.R. China}
\affiliation{School of Astronomy, School of Physical Sciences,
University of Science and Technology of China, Hefei 230026, P.R. China}

\begin{abstract}

We use Big Bang Nucleosynthesis (BBN) data in order to impose constraints on 
higher-order modified gravity, and in particular on:  (i) $f(G)$ Gauss-Bonnet 
gravity, and $f(P)$ cubic
gravities,  arising respectively through the use of the quadratic-curvature 
Gauss-Bonnet $G$ term, and the cubic-curvature combination, (ii) 
string-inspired 
quadratic Gauss-Bonnet  gravity coupled to the dilaton field,  
(iii) models with string-inspired quartic curvature corrections,  and 
(iv) running 
vacuum models. We perform a detailed investigation of the BBN epoch and we 
calculate the deviations of the freeze-out temperature $T_f$ in  comparison  to 
$\Lambda$CDM paradigm. We then use the observational  bound on $ 
\left|\frac{\delta {T}_f}{{T}_f}\right|$ in order to extract constraints on the 
involved parameters of various models. We find that all models can satisfy the 
BBN constraints and thus they constitute viable cosmological scenarios, since  
they can additionally
account for the dark energy sector and the late-time acceleration, in a 
quantitative manner, without spoiling the formation of light elements during 
the 
BBN epoch. Nevertheless, the obtained constraints on the relevant model 
parameters are quite strong.

\end{abstract}

  
\maketitle

\section{Introduction}

Modified gravity is one of the two main ways that are being followed in order 
to  explain the early and late accelerated phases of universe expansion 
\cite{Nojiri:2010wj,Clifton:2011jh,CANTATA:2021ktz,Ishak:2018his}, with the 
other one being the introduction of inflaton or/and dark energy sectors 
\cite{Copeland:2006wr,Cai:2009zp}.
Amongst the various classes of gravitational 
modifications that can fulfill the above cosmological motivation, 
theories that incorporate  higher-order corrections to the 
Einstein-Hilbert Lagrangian  have an additional motivation, namely 
the potential for improving the 
renormalizability of General Relativity \cite{Stelle:1976gc,Addazi:2021xuf}. 
Such theories 
 may naturally arise as (ghost-free) low-energy effective field-theory limits 
of 
String Theory \cite{Gross:1986mw}  and include Einstein gravity in the 
lowest-order in a derivative expansion. A particularly interesting
sub-class of such ghost-free higher-derivative theories that   
 are equivalent to General Relativity at the linearized level in 
the vacuum, with only a transverse and massless propagating graviton, are the 
 $($Lovelock$)$ theories \cite{Lovelock:1971yv}. 
 The most general, ghost-free covariant gravitational action in a Minkowski 
vacuum (that is, up to and including quadratic-order terms in fluctuations 
$h_{\mu\nu}$  of the graviton field $g_{\mu\nu}$ in the expansion 
$g_{\mu\nu}=\eta_{\mu\nu} + h_{\mu\nu}$, with $\eta_{\mu\nu}$ the Minkowski 
metric), which also involves higher-curvature as well as non-local terms with 
improved Ultraviolet behaviour (UV), but recovers Einstein's General Relativity 
in the Infrared (IR), has been given in~\cite{Biswas:2011ar}. 
 
The construction of higher-order gravities  is based on the addition
of extra terms in the Einstein-Hilbert Lagrangian, such as 
  in $f(R)$ gravity 
\cite{Starobinsky:1980te,Capozziello:2002rd,DeFelice:2010aj},
in Lovelock
gravity \cite{Lovelock:1971yv, Deruelle:1989fj}, in Weyl gravity
\cite{Mannheim:1988dj, Flanagan:2006ra}, in Galileon theory
\cite{Nicolis:2008in, Deffayet:2009wt}, etc. Restricting to 
quadratic-in-curvature
corrections a 
well-studied class 
is obtained by using functions of the Gauss-Bonnet combination, resulting to 
  the  $f(G)$
gravity \cite{Nojiri:2005jg}, which proves to have interesting 
cosmological phenomenology 
\cite{DeFelice:2008wz,DeFelice:2009aj,Bamba:2009uf,Elizalde:2010jx,
DeFelice:2010sh, delaCruzDombriz:2011wn,Makarenko:2012gm,Zhao:2012vta, 
Kofinas:2014daa,Chattopadhyay:2014xja,Saridakis:2017rdo,
Zhong:2018tqn,Lee:2020upe,Bajardi:2020osh,Shamir:2020ckh}. 
Similarly, using cubic terms one may construct   the particular cubic curvature 
invariance  $P$, which is a 
combination that is   neither 
topological nor
trivial in four dimensions and when used as a Lagrangian  leads  to a 
spectrum identical to that of 
General Relativity \cite{Bueno:2016xff}. Cubic and  
 $f(P)$ gravity have been also showed to lead to interesting cosmological 
\cite{Erices:2019mkd, 
Feng:2017tev,Poshteh:2018wqy,Marciu:2020ysf,Quiros:2020uhr,
Jimenez:2020gbw,
Nascimento:2020jwx}
and 
black-hole
applications 
\cite{ Hennigar:2016gkm,Bueno:2016lrh,Dykaar:2017mba, 
Hennigar:2018hza,Pavluchenko:2018jkm,Cano:2019ozf,
Burger:2019wkq, Konoplya:2020jgt}.

Additionally, in the context of microscopic string theory 
models~\cite{string,string2}, 
one faces situations where Gauss-Bonnet higher-curvature combinations couple to 
non-trivial dilaton fields, which also lead to 
interesting black hole solutions with secondary scalar hair~\cite{kanti} or 
modified cosmologies  with 
dilatons~\cite{GBsccosm,GBsccosm2,sami,lahanas,lahanas2,lahanas3,lahanas4a,lahanas4,
plionis}.
Finally, another class of such modification is   the  ``running vacuum 
models''  \cite{rvm1,rvm2,rvm3,rvmfoss,solaqft}, according to 
which the vacuum energy density   is a function of even powers of the Hubble 
parameter   and its derivative.   Such scenarios are shown to have 
interesting cosmology and phenomenology 
\cite{rvmpheno,rvmpheno2,rvmpheno2b,rvmevol,rvmevol2,rvmphenot,solayu,
rvmphenot2,rvmphenot3} and can 
additionally accept a microscopic string-inspired origin \cite{ms1,ms2,bms}.

An important and  necessary test of every modified gravity is the confrontation 
with cosmological observations, since such a confrontation   provides 
information on the involved unknown functions, as well as the allowed regions 
of the model parameters. Although, investigations related to late-time 
cosmological data~\cite{Planck} have been performed in some detail in the case 
of 
higher-order gravities \cite{DeFelice:2008wz,Elizalde:2010jx,Lee:2020upe}, the 
use of early-time, and in particular, of  
 Big Bang 
Nucleosynthesis (BBN) considerations has not been done as yet. Hence, in the 
present 
work we address this crucial issue,
namely we impose constrains on $f(G)$, Gauss-Bonnet-Dilaton  
and $f(P)$ gravities, and combinations thereof, as well as on running vacuum 
scenarios, through BBN analysis.

The plan of the article is the following: In Section \ref{Models} we 
  briefly present higher-order gravity, and in particular  $f(G)$, 
Gauss-Bonnet-Dilaton     $f(P)$ 
gravity,   as well as string-inspired models with quartic curvature 
corrections,    and running vacuum models, and we apply them in a 
cosmological 
framework. In Section 
\ref{BBNconstraints}, after a brief introduction to the basics of BBN, we 
examine in detail the BBN constraints on various specific models, extracting 
the bounds on the involved model parameters.  
Finally, Section \ref{Conclusions} is devoted to the Conclusions.

\section{ Higher-order gravity and cosmology}
\label{Models}

In this section we present higher-order gravity and we apply
it in a cosmological framework. As we mentioned in the Introduction, such 
theories are obtained through the addition of higher-order terms, that are 
constructed by contractions of
 Riemann tensors, in 
the Einstein-Hilbert Lagrangian  
\cite{Lovelock:1971yv}. Throughout the work we consider the flat
homogeneous and isotropic Friedmann-Robertson-Walker (FRW)   geometry with
  metric
\begin{equation}
\label{FRWmetric}
ds^{2}=-dt^{2}+a^{2}(t)\delta_{ij}dx^{i}dx^{j}\,,
\end{equation}
where $a(t)$ is the scale factor.
In the following subsections we examine the 
quadratic and cubic cases separately.

\subsection{$f(G)$ gravity and cosmology }

Let us first consider quadratic terms in the Riemann tensor. The corresponding 
combination is the Gauss-Bonnet one, given as\footnote{Our notation and conventions throughout this work are: signature of metric $(-, +,+,+ )$, Riemann Curvature tensor
$R^\lambda_{\,\,\,\,\mu \nu \sigma} = \partial_\nu \, \Gamma^\lambda_{\,\,\mu\sigma} + \Gamma^\rho_{\,\, \mu\sigma} \, \Gamma^\lambda_{\,\, \rho\nu} - (\nu \leftrightarrow \sigma)$, Ricci tensor $R_{\mu\nu} = R^\lambda_{\,\,\,\,\mu \lambda \nu}$, and Ricci scalar $R = R_{\mu\nu}g^{\mu\nu}$. We also work in units $\hbar=c=1$.}

\begin{equation}\label{GBinv}
G=R^2-4{R}_{\mu\nu}{R}^{\mu\nu}+{R}_{\mu\nu\rho\sigma}
{R}^{\mu\nu\rho\sigma}.
\end{equation}
Although this term is topological in four dimensions and thus it cannot lead 
to any corrections in the field equations,  the extended action
\begin{equation}\label{30}
S=\int d^4x 
\sqrt{-g}\left[\frac{M_P^2}{2}R+f\left(G\right)\right],
\end{equation}
with $M_{P}\equiv 1/\sqrt{8\pi \rm G_N} = 2.4 \times 10^{18}$~GeV the (reduced) 
Planck mass, and $\rm G_N$ the gravitational 
constant, corresponds to a new gravitational 
modification, namely 
$f(G)$ gravity. Variation of the action with respect to the metric 
leads to 
 \begin{eqnarray}
&&
\!\!\!\!\!\!\!
M_{P}^{2}{G}^{\mu\nu} =\frac{1}{2}{g}^{\mu\nu}f(G)-2f^{\prime}
\left(G\right)R    {R}^{\mu\nu}+4f^{\prime}
\left(G\right){R}^\mu_\rho {R}^{\nu\rho}\nonumber
\\
&& \
-2f^{\prime}
\left(G\right) 
R^{\mu\rho\sigma\tau}
R_\nu^{\rho\sigma\tau}-4f^{\prime}
\left(G\right)
R^{\mu\rho\sigma\nu} {R}_{\rho\sigma}\nonumber
\\
&&
\
+2\nabla^{\mu}\nabla^{\nu}f^{
\prime}
\left(G\right) 
R-2
{g}^{\mu\nu}\nabla^2f^{\prime}
\left(G\right)R+4\nabla^2f^{\prime}
\left(G\right)
 {R}^{\mu\nu}
\nonumber\\
&&
\
-4\nabla_{\rho}\nabla^{\mu}f^{\prime}
\left(G\right)
{R}^{\nu\rho}-4\nabla_{\rho}\nabla^{\nu}f^{\prime}
\left(G\right)
{R}^{\mu\rho}\nonumber
\\
&&\
+4
 {g}^{\mu\nu}\nabla_{\rho}\nabla_{\sigma}f^{\prime}
\left(G\right)
 {R}^{\rho\sigma}-4\nabla_{\rho}\nabla_{\sigma}f^{\prime}
\left(G\right)
{R}^{\mu\rho\nu\sigma},
\end{eqnarray}
with  $f_G\equiv \partial 
f(G)/\partial G$.
Applying it  to a cosmological framework, namely to the metric 
(\ref{FRWmetric}), and considering additionally the matter and radiation 
perfect fluids,  we find the Friedmann equations
\begin{eqnarray}
\label{FRWFR1}
&&3 M_{P}^2 H^2= \rho_m+\rho_r+\rho_{DE} \\
&&-2 M_{P}^2 \dot{H}= \rho_m+p_m+\rho_r+p_r+\rho_{DE}+p_{DE} ,\ \ \ \ \ 
\ 
\label{FRWFR2}
\end{eqnarray}
with $\rho_m$ and $p_m$   respectively the energy density and pressure of the 
matter   fluid, $\rho_r$ and $p_r$ the corresponding quantities for 
radiation 
sector,  and where we have introduced the  corresponding quantities of the  
effective dark 
energy 
sector as
 \begin{eqnarray}
&&
\!\!\!\!\!\!\!\!\!\!\!\!\!\!\!\!\!\!\!
\rho_{DE}\equiv   
\frac{1}{2}\left[-f\left(G\right)+24H^2\left(H^2+\dot{H}
\right)f^{\prime}\left(G\right)\right.\nonumber
\\
&&\left.-24^2H^4\left(2\dot{H}^2+H\ddot{H}+4H^2\dot{H}
\right)f^{\prime\prime}\left(G\right)\right] ,
\label{FRWrhoDE1}
\end{eqnarray}
\begin{eqnarray}
&&
\!\!\!\!\!\!\!\!\!\!\!\!\!\! 
p_{DE}\equiv
   f(G)-24H^2\left(H^2+\dot{H}\right)f^{\prime}
\left(G\right)\nonumber\\
&&
\ \ \,
+8(24)^2\left(2\dot{H}^2+H\ddot{H}+4H^2\dot{H}\right)^2f^{\prime\prime\prime}
\left(G\right)\nonumber\\
&&\ \ \,
+
192H^2\left(6\dot{H}^3+8H\dot{H}\ddot{H}+
24\dot{H}
^2H^2\right.
\nonumber\\
&&\left.\ \ \ \ \ \ \ \ \ \ \ \ \ \ \ +6H^3\ddot{H}+8H^4\dot{H}\right.
\left.+H^2\ddot{H}\right)f^{\prime\prime}
\left(G\right),
\label{FRWpDE1}
\end{eqnarray}
where primes denote differentiation with respect to the argument.
Note that in FRW metric the Gauss-Bonnet combination becomes  
 \begin{eqnarray}\label{GBfrw}
G=24H^2\big(H^2+\dot{H}\big),
\end{eqnarray}
which has  indeed squared powers comparing to the Ricci scalar 
$R=6(2H^2+\dot{H})$.

\subsection{Gauss-Bonnet-Dilaton Gravity}

An interesting situation, which stems directly from microscopic string theory 
models~\cite{string,string2}, is the modified gravity with quadratic curvature terms, 
which are coupled to non-trivial (dimensionless) dilatons $\Phi$, with potential $V(\Phi)$~\cite{kanti}. The effective 
action in this case  reads
 \begin{equation}\label{GBdil}
\int d^4x \sqrt{-g} \,M_{P}^2\, \Bigg[\frac{R}{2} - \frac{1}{4}\,  
\partial_\mu \Phi \, \partial^\mu \Phi + c_1 e^\Phi \, G  - V(\Phi)\Bigg],
\end{equation}
where the four-dimensionsl Gauss-Bonnet invariant $G$ is given in \eqref{GBinv}.
In string theory models the coefficient $c_1$ is given by
\begin{align}\label{c1g}
c_1 = \frac{\alpha^\prime}{8\,g_s^{(0)2}} , 
\end{align}
where $g_s^{(0)}$ is the string 
coupling and $\alpha^\prime =1/M_s^2$ is the Regge slope of the string, with 
$M_s$ the string scale, which is in general different from the (reduced) Planck 
mass $M_{P}$, entering the (3+1)-dimensional Einstein-Hilbert part of the low-energy string effective action \eqref{GBdil}. 
 
Einstein's and dilaton equations of motion in the presence of 
dilaton-Gauss-Bonnet coupling, stemming from \eqref{GBdil}, read~\cite{kanti}: 
\begin{eqnarray}
\label{GBeqs}
&&R_{\mu\nu} - \frac{1}{2} g_{\mu\nu} R  =  \frac{1}{2} \partial_\mu \Phi 
\partial_\nu \Phi - \nonumber \\
&&
\ \ \ \ \ \ \,
-\frac{1}{4}g_{\mu\nu}
\Big[(\partial_\alpha \Phi)^2
+ 4\, V(\Phi)\Big]
- \alpha^\prime \, \mathcal K_{\mu\nu}, \nonumber \\ 
&&
\mathcal K_{\mu\nu}   = 2\, (g_{\mu\rho}\, g_{\nu\lambda}   + g_{\mu\lambda}\, 
g_{\nu\rho}) \, \varepsilon^{\sigma\lambda\alpha\beta} D_\gamma \Big(\widetilde 
R^{\rho\gamma}_{\quad \alpha\beta}  \, \partial_\sigma f \Big)~,\nonumber \\  
&&
\partial_\mu (\sqrt{-g}\, \partial^\mu \Phi)  = - 2\, \sqrt{-g} \, c_1\, e^\Phi 
\, G - 2\, \frac{\delta V(\Phi)}{\delta \Phi}~,
 \end{eqnarray}
with $   f \equiv 
\frac{c_1}{\alpha^\prime}\, \, e^\Phi$. In the above equations $D_\mu$ 
denotes a gravitational covariant derivative, $\widetilde 
R^{\mu\nu}_{\quad\alpha\beta} = \frac{1}{2} \, \varepsilon^{\mu\nu\rho\sigma}\, 
R_{\rho\sigma\alpha\beta}$ 
is the dual Riemann tensor.  The gravitationally covariant Levi-Civita tensor density $\varepsilon_{\mu\nu\rho\sigma}$ is defined as $\varepsilon_{\mu\nu\rho\sigma} = \sqrt{-g} 
\,\epsilon_{\mu\nu\rho\sigma}$, with $\epsilon_{\mu\nu\rho\sigma} $ the flat-Mikowski-spacetime totally antisymmetric 
Levi-Civita symbol. The contraviariant tensor density $\varepsilon^{\mu\nu\rho\sigma}$  is defined accordingly by raising the indices with the appropriate curved metric tensor.
 The conserved stress tensor of the theory reads
\begin{equation}
\label{GBstress}
\mathcal T_{\mu\nu} = \frac{1}{2}\partial_\mu \Phi \, \partial_\nu  \Phi - 
\frac{1}{4} g_{\mu\nu}\Big[ (\partial_\alpha \Phi)^2  + 4\, V(\Phi)\Big] -
\alpha^\prime \, \mathcal K_{\mu\nu}~, 
\end{equation}
with 
\begin{align}
D_\mu \mathcal T^{\mu\nu} = 0.
\end{align}

Applying the above general field equations in the case of the flat FRW metric 
(\ref{FRWmetric}) we obtain the Friedmann equations 
\begin{eqnarray}
\label{a}
&&
\!\!\!\!\!\!\!\!\!\!\!\!
3H^2=  \frac{1}{4} \dot{\Phi}^2+V\left(\Phi\right)-24 c_1 
e^\Phi \dot{\Phi}H^3   +
M_P^{-2}\!\left(\rho_m\!+\!\rho_r\right)\!,\\
&&\!\!\!\!\!\!\!\!\!\!\!\!
2\dot{H}+3H^2=- \left[\frac{1}{4}\dot{\Phi}
^2-V\left(\Phi\right)+16 
c_1 e^\Phi\dot{\Phi} H(\dot{H}+H^2)\right.\nonumber \\
&&
\ \ \ \ \ \ \ \ \ \ \  
\left.+8 c_1 
e^\Phi    ( \ddot{\Phi}\!+\! \dot{\Phi}^2)H^2+ 
M_P^{-2}\!\left(p_m\!+\!p_r\right)\right]\!,
\end{eqnarray}
as well as the   dilaton equation
\be 
 \ddot{\Phi}+3H\dot{\Phi} 
+2V^{\prime}\left(\Phi\right)-48c_1 
e^\Phi H^2(\dot{H}+H^2)=0.
\label{dilatoneq}
\ee
Hence, we can identify the effective dark energy density   from (\ref{a}) as
\be
\rho_{DE}\equiv M_{P}^2\left[\frac{1}{4} 
\dot{\Phi}^2+V\left(\Phi\right)-24 c_1 
e^\Phi \dot{\Phi}H^3\right],
\label{rhodephi}
\ee
and the dark energy pressure as
\begin{eqnarray}
&&p_{DE}\equiv M_{P}^2\Big[\frac{1}{4}\dot{\Phi}
^2-V\left(\Phi\right)+16 
c_1 e^\Phi\dot{\Phi} H(\dot{H}+H^2)  
\nonumber\\
&&
\ \ \ \ \ \ \ \ \ \ \ \ \ \ \ \
 +8 c_1 
e^\Phi( \ddot{\Phi}+ \dot{\Phi}^2)H^2\Big].
\end{eqnarray}

\subsection{$f(P)$ gravity and cosmology}

We now proceed to the investigation of cubic terms. A general such combination 
is written as  \cite{Lovelock:1971yv}
\begin{eqnarray}\label{P}
&&
\!\!\!\!\!\!\!\!\!\!\!\!
P=\b_1 
{{{R_{\m}}^{\r}}_{\n}}^{\s}{{{R_{\r}}^{\g}}_{\s}}^{\d}{{{R_{\g}}^{\m}}_{\d}}^{\n
}+\b_2 
R_{\m\n}^{\r\s}R_{\r\s}^{\g\d}R^{\m\n}_{\g\d}
\nonumber\\
&&
+\b_3 
R^{\s\g}R_{\m\n\r\s}{R^{\m\n\r}}_{\g}+\b_4 
R R_{\m\n\r\s}R^{\m\n\r\s}\nonumber\\
&&
+\b_5 R_{\m\n\r\s}R^{\m\r}R^{\n\s}+\b_6 
R^{\n}_{\m}R^{\r}_{\n}R^{\m}_{\r}\nonumber\\
&&+\b_7 
R_{\m\n}R^{\m\n}R+\b_8 R^3.
\end{eqnarray}
Hence, using it as an argument of an arbitrary function we can construct 
the action of $f(P)$ gravity as \cite{Erices:2019mkd}
\begin{equation}\label{30b}
\int d^4x 
\sqrt{-g}\left[\frac{M_{P}^2}{2}R+f\left(P\right)\right].
\end{equation}
The above cubic combination possesses many coupling parameters. Nevertheless, 
we can significantly reduce their number by 
 requiring that in the case of simple cubic theory (i.e. with $f(P)=P$)
the resulting theory possesses a
spectrum identical to that of general relativity, that this combination  is 
neither topological nor
trivial in four dimensions, and that its definition is independent of the 
dimensions \cite{Bueno:2016xff}. Focusing additionally on FRW geometry we 
finally find that  \citep{Erices:2019mkd}  
\begin{equation}\label{24}
P=6\tilde{\beta}H^4\left(2H^2+3\dot{H}\right),
\end{equation}
which has only one free parameter. As expected is cubic in terms 
comparing to the Ricci scalar.

The two Friedmann equations of $f(P)$ gravity in the case of FRW geometry take 
the standard form (\ref{FRWFR1}),(\ref{FRWFR2}), however now the energy density 
and pressure of the effective dark energy fluid are written as  
\begin{eqnarray}
\label{rhofP}
&&
\!\!\!\!\!\!\!\!\!\!\!\!\!\!\!
\rho_{DE}   
 \equiv-f(P)-18\tilde{\beta} 
H^4(H\partial_t-H^2-\dot{H})f'(P),\\
&&\!\!\!\!\!\!\!\!\!\!\!\!\!\!\!
p_{DE} 
 \equiv f(P)+6\tilde{\beta}
H^3\left[H\partial_t^2+2(H^2+2\dot{H})\partial_t\right. \nonumber \\
&&\left.
\ \ \ \ \ \ \ \ \ \ \ \ \ \ \ \ \ \ \ \ \ 
-3H^3-5H\dot{H}\right]f'(P).
\label{pfP}
\end{eqnarray}

\subsection{String-inspired quartic curvature corrections \label{qcc}}

  We next proceed to discuss models inspired from string theory, which 
involve 
quartic curvature corrections~\cite{sami}. In the notation of that work  the 
effective 
low-energy string action is given in the form:
\begin{eqnarray}
\label{actionsami}
{\cal S} =\, \int d^D x \sqrt{-g}\left[R+
{\cal L}_{c}+\ldots\right]\,,
\end{eqnarray}
with $R$   the scalar curvature, and where for simplicity we use units where 
$M_P/2=1$. In the above expression   ${\cal L}_{c}$ denote the 
the string-inspired correction terms given by 
\begin{eqnarray}
{\cal L}_{c}=c_1\alpha^\prime e^{- 2 \Phi}
{\cal L}_2 + c_2 \alpha^{\prime 2} e^{- 4 \Phi}{\cal L}_3 + c_3
\alpha^{\prime 3} e^{- 6 \Phi}{\cal L}_4 \,,
\end{eqnarray}
where $\alpha'$ is the string Regge slope, serving as an expansion parameter, 
$\Phi$ is the 
dilaton field, and
\begin{eqnarray} 
& & {\cal L}_2 =\Omega_2\,, \\
& & {\cal L}_3 = 2 \Omega_3 +
R^{\mu\nu}_{\alpha \beta}
R^{\alpha\beta}_{\lambda\rho}
R^{\lambda\rho}_{\mu\nu}\,, \\
& & {\cal L}_4= {\cal L}_{41} -\delta_{H} {\cal L}_{42}
-\frac{\delta_{B}}{2}{\cal L}_{43}\,,
\end{eqnarray}
with
\begin{eqnarray} 
\hspace*{-1.0em}\Omega_2 &=& R^2-4R_{\mu \nu}R^{\mu \nu}+
R_{\mu \nu \alpha \beta}R^{\mu \nu \alpha \beta}, \\
\hspace*{-1.0em} \Omega_3 &=& R^{\mu\nu}_{\alpha \beta}
R^{\alpha\beta}_{\lambda\rho}R^{\lambda\rho}_{\mu\nu}
-2R^{\mu\nu}_{\alpha\beta}R_\nu^{\lambda\beta\rho}
R^\alpha_{\rho\mu\lambda} \nonumber \\
& & +{3\over 4} R R_{\mu\nu\alpha\beta}^2 + 
6  R^{\mu\nu\alpha\beta}R_{\alpha\mu}R_{\beta\nu} \nonumber \\
& &+ 4R^{\mu\nu}R_{\nu\alpha}R^\alpha_{\phantom{\alpha}\mu} - 6
R R_{\alpha\beta}^2 + \frac{R^3}{4},\\
\hspace*{-1.0em}  {\cal L}_{41} &=& \zeta(3)
R_{\mu\nu\rho\sigma}R^{\alpha\nu\rho\beta}\left(
R^{\mu\gamma}_{\delta\beta}
R_{\alpha\gamma}^{\delta\sigma}
- 2  R^{\mu\gamma}_{\delta\alpha}
R_{\beta\gamma}^{\delta\sigma} \right), \\
\hspace*{-1.0em} {\cal L}_{42} &=& {1\over 8} \left(
R_{\mu\nu\alpha\beta} R^{\mu\nu\alpha\beta}\right)^2
 +{1\over 4}  R_{\mu\nu}^{\gamma\delta}
R_{\gamma\delta}^{\rho\sigma}
R_{\rho\sigma}^{\alpha\beta}
R_{\alpha\beta}^{\mu\nu} \nonumber \\
& &- {1\over 2} R_{\mu\nu}^{\alpha\beta}
R_{\alpha\beta}^{\rho\sigma}
R^\mu_{\sigma\gamma\delta}
R_\rho^{\nu\gamma\delta}
- R_{\mu\nu}^{\alpha\beta}
R_{\alpha\beta}^{\rho\nu}
R_{\rho\sigma}^{\gamma\delta}
R_{\gamma\delta}^{\mu\sigma}, \\
\hspace*{-1.0em}  {\cal L}_{43} &=& \left(
R_{\mu\nu\alpha\beta}R^{\mu\nu\alpha\beta}\right)^2
- 10  R_{\mu\nu\alpha\beta}
R^{\mu\nu\alpha\sigma}
R_{\sigma\gamma\delta\rho}
R^{\beta\gamma\delta\rho} \nonumber \\
& &- R_{\mu\nu\alpha\beta}
R^{\mu\nu\rho}_{\sigma}
R^{\beta\sigma\gamma\delta}
R_{\delta\gamma\rho}^{\alpha}\,.            
\end{eqnarray}
The coefficient  $\delta_{H(B)}$ is equal to 1 for the case of the 
heterotic (bosonic) string theory and zero otherwise. 
The Gauss-bonnet term, $\Omega_{2}$, as well as  the Euler density, $\Omega_{3}$, 
do not contribute to the background equation of motion for $D=4$, 
unless the dilaton is dynamically evolving, which we shall not consider here.
The values of the coefficients $(c_1,\ c_2,\ c_3)$  depend
on the kind of the underlying string theories~\cite{sami}. In particular, we 
have
$(c_{1}, c_{2}, c_{3})=(0, 0, 1/8), (1/8, 0, 1/8), (1/4, 1/48, 1/8)$
for type II, heterotic, and bosonic strings, respectively. 

Following \cite{sami}, we consider a spatially flat (3+1)-dimensional 
Friedmann-Robertson-Walker metric
with a lapse function $N(t)$, namely $
ds^2=-N(t)^2dt^2+a(t)^2\sum_{i=1}^{3}\,
(dx^i)^2$,
and varying the effective action \eqref{actionsami} with respect to $N$ 
yields 
\begin{eqnarray}
\label{Hrhoc}
6H^2=\rho_c+\rho_m\,,
\end{eqnarray}
where  
\begin{eqnarray}
\label{rhoc}
\rho_c \equiv 
\frac{{\rm d}}{{\rm d}t}\left(\frac{\partial {\cal L}_c}
{\partial  \dot{N}}\right)+3H\frac{\partial {\cal L}_c}
{\partial  \dot{N}}-\frac{\partial {\cal L}_c}{\partial N}
-{\cal L}_{c} \biggr|_{N=1} \,.
\end{eqnarray}
The quantity $\rho_{m}$ denotes the energy density of the cosmic fluid, which in \cite{sami}
was assumed to be of 
barotropic type, with an equation of state
$
w_m = p_m/\rho_m$
satisfying the conservation equation
\begin{eqnarray}
\label{rhom}
\dot{\rho}_{m}+3H(1+w_{m})\rho_{m}=0\,.
\end{eqnarray}
{}From Eq.~(\ref{rhoc}) we find that 
for bosonic strings, 
the energy density $\rho_c$ is given by~\cite{sami}
\begin{eqnarray}
\label{corre2}
\rho_c &=&
A (5H^6+2I^3-6HIJ) +B \{[-21\zeta(3)+210]H^8
\nonumber \\
& &-[3\zeta(3)-90]I^4-[12\zeta(3)+48]H^4I^2
\nonumber \\
&& +
[4\zeta(3)+120]H^2I^3 -[24\zeta(3)-96]H^6I\nonumber \\
&&
+J \{[8\zeta(3)-32]H^5+[12\zeta(3)-360]HI^2
\nonumber \\
&&+
24\zeta(3)H^3I \}\}\,,
\end{eqnarray}
where $I \equiv H^2+\dot{H}$, 
$J=\ddot{H}+3H\dot{H}+H^3$, 
$A=24c_2\alpha'^2e^{-4\phi}$ and $B=6c_3 \alpha'^3 e^{-6\phi}$.

For type II \& heterotic strings we have \cite{sami} 
\begin{eqnarray}
\label{corre}
\rho_c &=& B [a_8H^8+a_c I^4+a_4H^4I^2+a_2H^2I^3+a_6H^6I
\nonumber \\
& & -J(a_5H^5+a_1HI^2+a_3H^3I)]\,,    
\end{eqnarray}
with $B=6c_3 \alpha'^3 e^{-6\phi}\zeta(3)$, 
$a_8 =-21, a_c=-3, a_4=-12, a_2=4, a_6=-24, a_5=-8,
a_1=-12, a_3=-24$ for type II string, and $B=6c_3 \alpha'^3 
e^{-6\phi}$, $a_8=-21\zeta(3)+35, a_c=-3\zeta(3)+15, a_4=-12\zeta(3)-6,
a_2= 4\zeta(3)+20, a_6=-24\zeta(3)+12, a_5=-8\zeta(3)+4,
a_1=-12\zeta(3)+60, a_3=-24\zeta(3)$ for the heterotic string.

Let us make a comment here on the relation  of this model with the case of 
cubic corrections of the previous sections. In the case of cubic and $f(P)$ 
gravity the coefficient parameters in  (\ref{P}) have been chosen in order for 
$P$ not to have higher than second-order derivatives, while in the present 
model this condition is not required (one can see that (\ref{corre2}) and 
(\ref{corre}) contain higher derivatives). Hence, even in the limit $c_3=0$ the 
scenario with string-inspired quartic curvature corrections does not recover 
  the cubic scenario.

Finally, in \cite{sami} the authors considered the case of phantom energy, with 
$w_m < -1$. The analysis showed, for instance, 
that for the case of type II string theories, universes with 
$w_m \lesssim -1.2$ approach the Big Crunch singularity, while for $-1.2 \lesssim w_m < -1$, they tend to approach the 
Big Rip singularity. 

In this work  we are interested in    examining these 
cases from the point of view of BBN constraints, as we will see in subsection 
\ref{sec:qcc} below.    

\subsection{Running vacuum   cosmology \label{RVM}}

In this subsection we briefly review cosmology with running vacuum 
\cite{rvm1,rvm2,rvm3,rvmfoss,solaqft,rvmpheno,rvmpheno2,rvmpheno2b,rvmevol,
rvmevol2,rvmphenot,solayu,rvmphenot2,rvmphenot3}. In such a scenario  
the vacuum energy density is expressed, as a result of general covariance, in 
terms of even integer powers of the Hubble parameter. We mention that possible 
dependence in terms of $\dot H$, during the brief epoch of BBN in which we are 
interested in, can be expressed   in terms of $H^2$  using the approximately 
constant value of the deceleration parameter  for that period, and hence we 
shall not consider it explicitly here (nonetheless, the current-era 
phenomenology of such  $\dot H$ terms has been examined with precision (for 
latest work see, e.g. \cite{rvmphenot}).

The corresponding running vacuum model 
(RVM) energy density, which plays the role of the dark energy, reads as
\begin{align}\label{rvmener}
\rho^{\rm \rm vac}_{\rm RVM} = 3 M_P^2 \Big(c_0 + \nu H^2 + \frac{\alpha 
}{H_I^2} \, H^4 + \dots \Big),
\end{align}
while the RVM equation of state is~\cite{rvm1,rvm2,rvm3,rvmfoss}
\begin{align}\label{rvmeos}
p_{\rm RVM}^{\rm vac} = - \rho^{\rm \rm vac}_{\rm RVM},
\end{align}
in the standard parametrisation  where $H_I$ is the inflationary scale, which 
according to Planck data~\cite{Planck} is of the order of $H_I \sim 10^{-5} \, 
M_P$. Furthermore, the coefficients $\nu, \alpha > 0$ are dimensionless, and 
the neglected terms in \eqref{rvmener} denote terms of  order $H^6$ and higher 
since they do not play a significant role  \cite{rvmevol,rvmevol2}.
The constant $c_0$ in \eqref{rvmener} cannot be determined from first 
principles, given that the model is based on an expansion in even powers of $H$ 
of the ``renormalisation-group-like scaling" $\frac{d \, \rho_{\rm RVM}^{\rm 
vac}}{d \, {\rm ln} H} = \sum_{n=1}^\infty c_n \, H^{2n} $. In cosmological 
terms  $c_0$ plays the role of the cosmological constant. In such a 
case, the term $\nu H^2$ yields observable deviations from $\Lambda$CDM 
scenario, and the current phenomenology requires $\nu = 
\mathcal O(10^{-3}) > 0$ \cite{rvmpheno,rvmpheno2b,rvmphenot3}. Moreover, 
RVM-like models can be used to alleviate  the tensions \cite{Abdalla:2022yfr} 
that characterise the current-era cosmological 
data~\cite{rvmpheno,rvmpheno2,rvmphenot,rvmphenot2} (for the phenomenology of 
more general models 
of $\Lambda (t)$-varying cosmologies, see \cite{tsiapi,tsiapi2}).

Microscopic models for RVM  are constructed within the context of 
string-inspired cosmological scenarios \cite{ms1,ms2,bms}. In such models, 
which involve gravitational anomalies (such as Chern-Simons terms) coupled to 
axion fields, the term $H^4$ term, which may drive dynamical inflation,   is 
induced by a CP-violating primordial gravitational-wave condensate of the 
Chern-Simons terms. 
In the post inflationary epoch  gravitational anomalies are 
assumed to be cancelled, and therefore the term  $H^4$ will be absent   during 
BBN. Nonetheless, in general, there may be other, non-geometric, reasons for 
having such $H^4$ terms, for instance as a result of integrating out interacting 
quantum fields of some mass scale $m$. 

In general, as discussed in \cite{ms1,nmtorsion,nmtorsion2} (see also 
\cite{houston,basilsugra} in the case of (dynamically broken) supergravity, 
which can be linked to early-universe running vacuum~\cite{ms1,nmtorsion2}), 
quantum graviton corrections around a 
background cosmological space-time might also lead to a renormalisation of the 
Newton constant, which becomes  ``running'' too, 
exhibiting a mild logarithmic dependence on the Hubble parameter. The precise 
form of such a running will depend on the underlying theory of quantum gravity.\footnote{The supergravity models 
of \cite{houston} involve logarithmic corrections of the one-loop induced 
de-Sitter cosmological constant $\Lambda >0$, which in a general relativity 
setting is proportional to the curvature scalar $R$, and thus proportional to 
$H^2$ in cosmological models. For our purposes we assume a mild cosmic time 
dependence of $H(t)$. In this sense,   the quantum graviton integration in such 
models leads to ${\rm ln}(R) R^n, n \in \mathbb Z^+$ terms in an effective 
action, which in a cosmological setting leads to energy densities of the form 
\cite{ms1,nmtorsion,nmtorsion2}. 
It should be stressed that the above computation in the model of \cite{houston} 
refers to  weakly-coupled graviton fluctuations about a (de Sitter) background 
in a fixed {\it gauge}. The issue of gauge invariance in a full quantum gravity 
theory is a complicated issue, and will not be discussed here. We refer the 
reader to \cite{batalin,batalin2} for more details on this important issue. We 
also notice that modified gravity models involving ${\rm ln}(R)$ terms have been 
considered in \cite{nojiri}, but from a different perspective than ours. 
Moreover, in our case the corrections have a smooth flat limit $R \to 0$.}   
In this respect, for our phenomenological purposes here, we may parametrize the 
effective RVM vacuum energy density during the post-inflationary eras, 
including the BBN epoch, as \cite{ms1,nmtorsion,nmtorsion2}:
 \begin{eqnarray}\label{logH2}
 &&\!\!\!\!\!\!\!\!\!\!\!\!\!\!\!\!\! 
 \rho_{\rm RVM}^{\rm vac}\equiv \rho_{DE}   \simeq  3 M_P^2 \Big\{c_0 + [ \nu + 
d_1\, {\rm ln}(M_P^{-2}H^2) ] \, H^2 \nonumber \\ 
&& \ \ \ \ \ \ \ \ \    +  \frac{\alpha}{H_I^2} [ 1 + d_2\, 
{\rm ln} (M_P^{-2}H^2) ] \, H^4 + \dots\Big\} ~,  
 \end{eqnarray}
where $c_0 > 0, \nu > 0, \alpha > 0$ and  $d_i, i=1,2$ are phenomenological parameters.

\section{Big Bang Nucleosynthesis  constraints } 
\label{BBNconstraints}

In this section we will investigate the Big Bang Nucleosynthesis (BBN) 
constraints on scenarios that are governed by higher-order modified gravity.
 BBN is realized during the radiation epoch 
\cite{Bernstein:1988ad,Kolb:1990vq,Olive:1999ij,Cyburt:2015mya}.  In the case 
of 
standard cosmology, i.e. in 
the case of Standard Model radiation in the framework of general relativity, 
during the BBN
the  first Friedmann equation is
approximated as
\begin{eqnarray}
 H^2\approx\frac{M_{P}^{-2}}{3} 
 \rho_r\equiv H_{GR}^2.
 \label{Frrad1}
\end{eqnarray}
Additionally, we know that 
 the energy density of relativistic particles  is
 \begin{eqnarray}
 \label{Temptime}
{\displaystyle \rho_r=\frac{\pi^2}{30}g_* {T}^4},
\end{eqnarray}
with $g_*\sim 10$  the effective number of degrees of freedom    
and ${T}$   the temperature. Hence, we obtain
\begin{eqnarray}
 H(T) \approx    \left(\frac{4\pi^3 
g_*}{45}\right)^{1/2}\frac{{T}^2}{M_{Pl} 
},
\label{HTemprel}
\end{eqnarray}
   where  $M_{Pl}= (8\pi)^{\frac{1}{2}} M_P = 1.22 \times 10^{19}$ GeV is the Planck mass.

Since the radiation conservation equation finally leads to 
a scale factor evolution of the form 
$a\sim
t^{1/2}$, we can finally   extract the 
expression
between temperature
and time, namely ${\displaystyle \frac{1}{t}\simeq 
\left(\frac{32\pi^3 
g_*}{90}\right)^{1/2}\frac{{T}^2}{M_{Pl}}
}$
(or
${T}(t)\simeq (t/\text{sec})^{-1/2} $~MeV).

During the BBN, the calculation of the neutron abundance 
arises from the protons-neutron
conversion rate \cite{Olive:1999ij,Cyburt:2015mya}
 \begin{equation}
 \lambda_{pn}({T})=\lambda_{(n+\nu_e\to p+e^-)}+\lambda_{(n+e^+\to p+{\bar
\nu}_e)}+\lambda_{(n\to p+e^- +
{\bar \nu}_e)}\,
 \end{equation}
and its inverse $\lambda_{np}({T})$, and therefore for the  total 
rate we have
$    \lambda_{tot}({T})=\lambda_{np}({T})+\lambda_{pn}({T})$.
 Assuming that the varius particles (neutrinos, electrons, photons) 
temperatures   are the same, and low enough in order to use 
the Boltzmann distribution instead of the
Fermi-Dirac one),  and neglecting  the electron mass  
 compared to the electron and neutrino energies, straightforward calculations 
lead to the expression 
\cite{Torres:1997sn,Lambiase:2005kb,Lambiase:2011zz,Capozziello:2017bxm,
Barrow:2020kug}
 \begin{equation}\label{Lambdafin}
    \lambda_{tot}({T}) =4 A\, {T}^3(4! {T}^2+2\times 3! {  Q}{T}+2!
{  Q}^2)\,,
 \end{equation}
where  ${
Q}=m_n-m_p=1.29 \times10^{-3}$GeV is   the   mass difference between neutron 
and proton 
and $A=1.02 \times 10^{-11}$~GeV$^{-4}$.

Let us now calculate the corresponding freeze-out temperature. This will arise 
from the 
comparison of the universe expansion rate $\frac{1}{H}$ with $ 
\lambda_{tot}\left(T\right)$. In particular, if $\frac{1}{H} \ll
\lambda_{tot}\left(T\right)$, namely if the expansion time is much smaller than
the interaction time we can consider    thermal equilibrium 
\cite{Bernstein:1988ad,Kolb:1990vq}. On the contrary, if $\frac{1}{H}
\gg\lambda_{tot}\left(T\right)$ then particles 
 do not have enough   time   to interact and therefore they 
 decouple. Thus,  the freeze-out 
temperature $T_{f}$, in which the decoupling takes place corresponds to 
 $H (T_{f})= \lambda_{tot}\left(T_{f}\right) \simeq c_q \,T_f^5$, with $c_q 
\equiv 4A \, 4! \simeq 9.8 \times 10^{-10} \, {\rm GeV}^{-4}$
\cite{Torres:1997sn,Lambiase:2005kb,Lambiase:2011zz,Capozziello:2017bxm,
Barrow:2020kug}.  
 Using (\ref{HTemprel}) and (\ref{Lambdafin}), the above requirement gives
\begin{equation}
T_{f}=\left(\frac{4\pi ^3 g_{*}}{45M_{Pl}^2c_{q}^2}\right)^{1/6}\sim 0.0006 
~{\rm 
GeV}.
\end{equation}

Now, in any modified cosmological scenario  one obtains    extra terms in 
the Friedmann equations. During the BBN era these extra contributions 
 need to be small, compared to the radiation sector of  
standard cosmology, in order not to spoil the observational facts.
In particular, from a general modified Friedmann equation of the form 
(\ref{FRWFR1}) we obtain
\begin{equation}
H=H_{GR}\sqrt{1+\frac{\rho_{DE}}{\rho_{r}}}=H_{GR}+\delta H,
\end{equation}
where $H_{GR}$ is the Hubble parameter of standard cosmology.  
Thus, we have
\begin{equation}
\delta H=\left(\sqrt{1+\frac{\rho_{DE}}{\rho_{r}}}-1\right)H_{GR}.
\end{equation}
This deviation  from standard cosmology, i.e form  $H_{GR}$, will lead to a 
deviation in the  freeze-out 
temperature  $\delta 
T_{f}$. Since   $H_{GR} = \lambda_{tot}\approx c_q\,T_f^5$, we easily find
\begin{equation}
\left(\sqrt{1+\frac{\rho_{DE}}{\rho_{r}}}-1\right)H_{GR}=5c_q\,T_{f}^4\delta 
T_{f},
\end{equation}
and finally 
\begin{equation}
\frac{\delta 
T_{f}}{T_{f}}\simeq\frac{\rho_{DE}}{\rho_{r}}\frac{H_{GR}}{10c_q\,T_{f}^5},
\label{finalexpress}
\end{equation}
 where we used that  $\rho_{DE}<<\rho_{r}$ during BBN. 
This theoretically calculated $\frac{\delta 
T_{f}}{T_{f}}$ should be compared with the observational bound
 \begin{equation}
 \label{deltaTbound}
    \left|\frac{\delta {T}_f}{{T}_f}\right| < 4.7 \times 10^{-4}\,,
 \end{equation}
 which is obtained from the observational estimations of
the baryon mass
fraction   
converted to ${}^4 He$ 
\cite{Coc:2003ce,Olive:1996zu,Izotov:1998mj,Fields:1998gv,Izotov:1999wa,
Kirkman:2003uv, Izotov:2003xn}.
 In the following subsections we use the above formalism, and in particular 
expression (\ref{finalexpress}), in order to impose constraints on $\rho_{DE}$ 
and thus on the underlying modified gravity, in   specific models.

\subsection{$f(G)$ Gravity }

In this section we apply the above formalism in order to extract the BBN 
constraints on $f(G)$ gravity. In particular, we will use the dark energy 
density expression (\ref{FRWrhoDE1}) that holds in the case of $f(G)$ 
cosmology. In order to proceed we need to examine specific $f(G)$ forms. We 
will focus on three well-studied models that are known to lead to interesting 
late-time cosmology. 

We stress at this point that, since we are interested in 
scenarios where the $f(G)$ term gives rise to the effective dark energy sector, 
we do not consider the case where  $f(G)$ is linear in $G$, since in this case 
there is no contribution from $G$ to the Friedmann equations, given its 
topological (total-derivative) nature in (3+1)-dimensional space times.

\subsubsection{$f(G)$ Model I }

\begin{figure}[ht] 
\centering
\includegraphics[angle=0,width=0.49\textwidth]{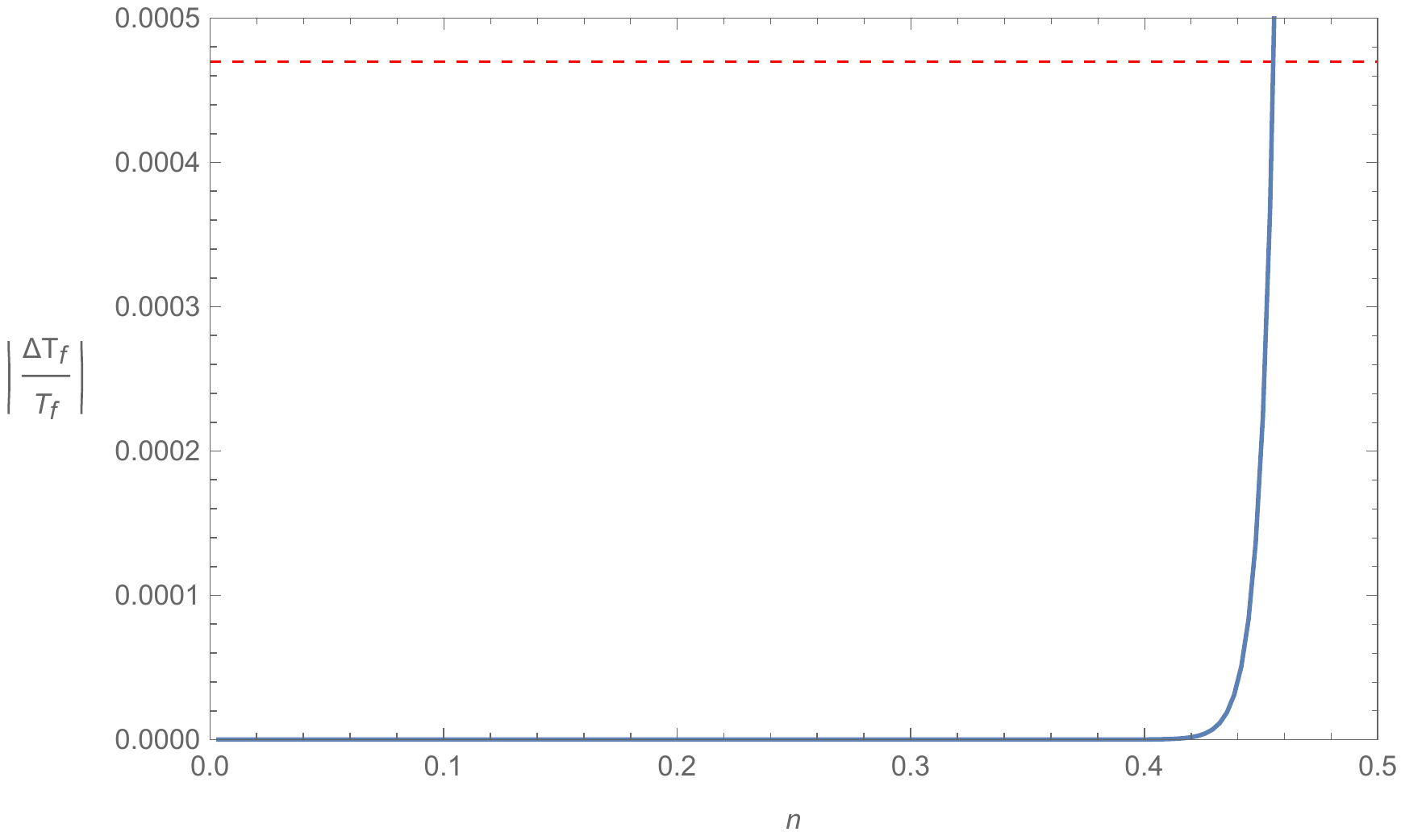}
\vspace{-0.2cm}\caption{{
\textit{ $\delta {  T}_f/{  T}_f$ from
(\ref{fG1finalexpr})
vs the model parameter $n$   (blue solid  curve), in the case of $f(G)$ Model I 
 of (\ref{fGmod1}), and 
the upper bound
for $\delta {  T}_f/{  T}_f$ from (\ref{deltaTbound}) (red dashed line). 
As 
we observe, as long as  
$n\neq1$  constraints from BBN require $n\lesssim 0.45$ (in the case $n=1$ the 
above model cannot account for 
dark energy due to the topological nature of $G$).}} }
\label{figfG1}
\end{figure}

As a first example, namely $f(G)$ Model I,  we consider the power-law model  
\cite{Nojiri:2005jg} where 
\begin{equation}\label{fGmod1}
f\left(G\right)=\alpha G^{n},
\end{equation}
with $n\neq1$. In this expression $n$ is the only free model parameter, since as 
long as  
$n\neq1$ then $\alpha$ can be 
expressed in terms of the present value of the Hubble parameter $H_0$ and the 
present value of the dark energy density parameter $\Omega_{DE0}\equiv 
\rho_{DE0}/(3M_P^{2}H_0^2)$ (in the case $n=1$ the above model cannot account 
for 
dark energy, in view of the aforementioned total derivative nature of the 
four-dimensional Gauss-Bonnet invariant). 
In particular, by applying (\ref{FRWrhoDE1}) 
at present we find
\be
\alpha=\frac{3 
H_{0}^2\Omega_{DE0}}{M_P^{-2}\left[\left(n-1\right)G_{0}^{n}
-n\left(n-1\right)\gamma_{0}G_{0}^{n-2}\right]},
\label{condmodel1}
\ee
where 
\begin{equation}
G_{0}=24H_{0}^2\left(H_{0}^2+\dot{H}_{0}\right),
\label{G0rel}
\end{equation}
and
\begin{equation}
\gamma_{0}=24^2H_{0}^4\left(2\dot{H}_{0}^2+H_{0}\ddot{H}_{0}+4{H}_{0}^2\dot{H}_{
0}\right).
\end{equation}
Inserting (\ref{fGmod1}) into (\ref{FRWrhoDE1}) and then into 
(\ref{finalexpress}) we finally obtain
\begin{eqnarray}
&&
\!\!\!\!\!\!\!\!\!\!\!\!\!\!\!
\frac{\delta 
T_{f}}{T_{f}}=-\Omega_{DE0}\left(\zeta\right)^{4n-1}\left(T_{f}\right)^{
8n-7}\nonumber\\
&&\cdot\left[\left(-1\right)^{n}+n(-1)^{n-1}+8n\left(n-1\right)\left(-1\right)^{
n-2}\right]\nonumber\\
&&
\cdot(H_{0})^{2-2n}\left(H_{0}^2+\dot{H}_{0}\right)^{
-n}\left[10c_q\left(n-1\right)\right]^{-1}\nonumber\\
&&\cdot
\left[\left(1\!-\!2n\right)\!\left(\dot{H}_{0}^2\!+\!2\dot{H}_{0}H_{0}
^2\right)- n\ddot
{H}_{0}H_{0}+
H_{0}^4\right]^{-1},
\label{fG1finalexpr}
\end{eqnarray}
with
\begin{align}\label{defzeta}
\zeta\equiv\left(\frac{4\pi^3g_{*}}{45}\right)^{\frac{1}{2}}M_{Pl.}^{-1}~.
\end{align}
In this expression we set~\cite{Planck} 
\begin{align}\label{deo}
\Omega_{DE0} \approx0.7, \quad 
 H_{0}=1.4\times 
10^{-42} ~{\rm GeV}, 
\end{align}
and   the derivatives of the Hubble 
function at present are calculated through 
 $\dot{H}_{0}=-H_{0}^2\left(1+q_{0}\right)$ and 
$\ddot{H}_{0}=H_{0}^3\left(j_{0}+3q_{0}+2\right)$
 with $q_{0}=-0.503$ the current decceleration parameter of the 
Universe~\cite{Planck}, and $j_{0}=1.011$ the current jerk parameter 
\cite{Visser:2003vq,Mamon:2018dxf}.
 In Fig. \ref{figfG1} we plot $\delta {  T}_f/{  T}_f$ appearing in 
(\ref{fG1finalexpr})
 vs the model parameter $n$, as well as the upper bound inferred from 
(\ref{deltaTbound}). As becomes evident from the figure,  
the expression (\ref{fG1finalexpr})   satisfies the bound  (\ref{deltaTbound}) 
for
$n\lesssim 0.45$.

 We stress here that in this work we desire to impose BBN constraints on 
higher-order modified gravities models that  can describe dark energy. Hence, 
concerning the present $f(G)$ model we require  the fulfillment of   condition 
(\ref{condmodel1}), which imposes a dependence of the model parameters $\alpha$ 
and $n$. That is why BBN  analysis leads to a so strong constraint on $n$. If 
we relax   condition (\ref{condmodel1}) then the BBN 
constraints can always be fulfilled for every $n$ by suitably constraining 
$\alpha$, and equivalently the BBN  constraints can always be fulfilled for 
every $\alpha$ by suitably constraining $n$. However, under  the condition 
(\ref{condmodel1}), namely under the requirement that the Gauss-Bonnet terms 
describe dark energy at the late Universe, then $n$ is constrained close to 
zero, in which case $f(G)$ correction becomes a constant and the scenario 
becomes $\Lambda$CDM.

\subsubsection{$f(G)$ Model II}

As a second concrete example we consider the $f(G)$ Model II,  namely 
\cite{DeFelice:2008wz}
 \begin{eqnarray}
&&
\!\!\!\!\!\!\!\!\!\!\!\!
f(G)=\lambda\frac{G}{\sqrt{G_{*}}}\arctan\left(\frac{G}{G_{*}}\right)-\frac{
\lambda}{2}\sqrt{G_{*}}\ln\left(1+\frac{G^2}{G_{*}^2}
\right)\nonumber\\
&&
\ \ \ 
-\alpha\lambda\sqrt {
G_{*}},
\label{fGmod2}
\end{eqnarray}
where   $\alpha$,  $\lambda>0$ and $G_{*}>0$ are the model parameters.
Applying (\ref{FRWrhoDE1}) at present time we can express 
 $\lambda$ as
  \begin{eqnarray}
&&  \!\!\!\!\!\!\!\!\!\!
\lambda= 3M_{P}^2H_{0}^2 \Omega_{DE0}
\Bigg\{\arctan\left(\frac{G_{0}}{G_{*}}\right)\sqrt{G_{0}}
\left(1-\sqrt{\frac{G_{0}}{G_{*}}}\right)
 \nonumber\\
&& +\sqrt{G_{*}}\left[\alpha+\frac{1}{2}
\ln\left(1+\frac{G_{0}^2}{G_{*}^2}
\right)-\frac{\gamma_{0}}{G_{0}^2+G_{*}^2}\right]\Bigg\}^{-1},
\end{eqnarray}
and thus we remain with only two free parameters $\alpha$  and 
$G_{*}$. 
Inserting (\ref{fGmod2}) into (\ref{FRWrhoDE1}) and then into 
(\ref{finalexpress}) we obtain
\begin{eqnarray}
&& 
\!\!\!\!\!\!\!\!\!\!\!\!\!\!\!\!
\frac{\delta 
T_{f}}{T_{f}}=\frac{1}{20}\Omega 
_{DE0}H_{0}^2\sqrt{G_{*}}
\left[c_q\,\zeta T_{f}^{7}\right]^{-1} 
\nonumber\\
&&
\ \,  
\cdot
\left\{
2\alpha+\ln\left[ 
1+\frac{\zeta^8T_{f}^{16}}{G_{*}^2}\right]
\right. \nonumber\\
&&\left.
\ \ \ \ \,     -
\frac{ 
9216\zeta^{8}T_{f}^{16}}{576\zeta^8 T_{f}^{16}+G_{*}^2}
\right\}\\ 
&&
\ \,    
\cdot
\Bigg\{\!\arctan\left(\frac{G_{0}}{G_{*}}\right)\sqrt{G_{0}}
\left(1-\sqrt{\frac{G_{0}}{G_{*}}}\right) \nonumber \\
&&\ \ \ \ \,      
+\sqrt{G_{*}}\left[\alpha\!+\!\frac{1}{2}\ln\!\left(1\!+\!\frac{G_{0}^2}{G_{*}^2
}
\right)\!-\!\frac{\gamma_{0}}{G_{0}^2\!+\!G_{*}^2}\right]\!\!\Bigg\}^{-1}\!\!,
\label{fG2finalexpr}
\end{eqnarray}
with $\zeta$ given in \eqref{defzeta}. 

Interestingly enough, expression
 (\ref{fG2finalexpr}) always satisfies the bound  (\ref{deltaTbound}) for the 
$\alpha$ regions that are needed in order to have  a stable de Sitter 
point  \cite{DeFelice:2008wz}
  (that is, $G_{*}^{\frac{1}{4}}\sim 10^{-42}$~GeV  and 
$\alpha\approx100$ and  $\lambda\ll1$). Hence, BBN 
cannot impose any constraints on this model beyond those already obtained from 
theoretical consistency. 

\subsubsection{$f(G)$ Model III}

We proceed to the investigation of the  $f(G)$ Model III, for which
\cite{DeFelice:2008wz} 
\begin{equation}
f_{2}(G)=\lambda\frac{G}{\sqrt{G_{*}}}\arctan\left(\frac{G}{G_{*}}
\right)-\alpha\lambda\sqrt{G_{*}}.
\end{equation}
Applying (\ref{FRWrhoDE1}) at present time we eliminate  
 $\lambda$, and, then, on inserting into (\ref{finalexpress}) we 
  find
\begin{eqnarray}
&&
\!\!\!\!\!\!\!\!\!\!\!\!\!
\frac{\delta T_{f}}{T_{f}}=\frac{1}{10}\Omega 
_{DE0}H_{0}^2\left[c_q\,\zeta T_{f}^7\right]^{-1
}
\nonumber
\\
&&\!\!\!\!\!\!\!\!\!\cdot
\left\{ 
\frac{\left(G_{0}^2+G_{*}^2\right)\left[
\left(1\!+\!\alpha\right)G_{0}^2+\alpha 
G_{*}^2)\right]-2\gamma_{0}G_{*}^2}{\left(G_{0}^2+G_{*}^2\right)^2}\right\}^{-1
}
\nonumber
\\
&&\!\!\!\!\!\!\!\!\!
\cdot\left\{ 
 -\frac{576\zeta^8T_{f}^{16} 
}{576\zeta^8T_{f}^{16}+G_{*}^2} 
-\alpha   +\frac{9216\zeta^8T_{f}^{16}G_{*}^2}{\left[
576\zeta^8 T_{f}^{16}+G_{*}^2\right]^2}\right\},
\label{fG3finalexpr}
\end{eqnarray}
with $G_0$ given in (\ref{G0rel}) and  
  $\zeta$   in \eqref{defzeta}. 
Similarly to the previous model, we find that 
expression
 (\ref{fG3finalexpr}) always   satisfies the bound  (\ref{deltaTbound}) for the 
parameter regions that are needed in order to have  a stable de Sitter 
point  \cite{DeFelice:2008wz}.
Thus, BBN cannot impose  stinger constraints on this model, i.e. it always 
satisfies BBN requirements.

\subsubsection{$f(G)$ Model IV }

As a last model we consider \cite{DeFelice:2008wz}
\begin{equation}
f(G)=\lambda\sqrt{G_{*}}\ln\left[\cosh\left(\frac{G}{G_{*}}\right)\right]
-\alpha\lambda\sqrt{G_{*}}.
\end{equation}
Applying (\ref{FRWrhoDE1}) at present time we eliminate  
 $\lambda$ and then inserting into (\ref{finalexpress}) we 
  find
\begin{eqnarray}
&&
\!\!\!\!\!\!\!\!\!\!\!\!\!\!\!\!
\frac{\delta T_{f}}{T_{f}}=\frac{1}{10}\Omega 
_{DE0}H_{0}^2\Big\{4608\zeta^8 \text{sech}(\xi 
^2)T_{f}^{16}  \nonumber
\\ \ \
&&
 +
[\ln(\cosh\xi)
-\alpha]G_{*}
^2 -24\zeta^4T_{f}^8G_{*}\tanh\xi\Big\}\nonumber
\\ \ \
&&\cdot\left[
c_q\zeta T_{f}^7\right]^{-1}
\nonumber\\
&&
\left\{\gamma_{0}
\text{sech}\left(\frac{G_{0}}{G_{*}}\right)^2  
+\left\{\ln\left[\cosh\left(\frac{
G_ {
0}}{G_{*}}\right)\right]-
\alpha\right\}G_{*}^2\right.\nonumber
\\
&& \ \  
\left.-G_{0}G_{*}\tanh\left(\frac{G_{0}}
{
G_{*}}\right)\right\}^{-1},
\end{eqnarray}
with
$\zeta$ defined in \eqref{defzeta} and 
$\xi=\frac{24\zeta^4T_{f}^8}{G_{*}}$.

Similarly to the previous models,  
expression
 (\ref{fG3finalexpr}) always   satisfies the bound  (\ref{deltaTbound}) for the 
parameter regions required in order to have  a stable de Sitter 
point  \cite{DeFelice:2008wz}, and therefore BBN   requirements are always 
fulfilled.

\subsection{Gauss-Bonnet-Dilaton Gravity}

The case of Gauss-Bonnet-Dilaton gravity is more complicated than $f(G)$ and 
$f(P)$ gravity, since one has the additional dilaton field $\Phi$, which 
follows its own evolution   (equation (\ref{dilatoneq})). Thus, in general one 
cannot extract any analytical expressions and one needs to solve numerically 
the whole system of Friedmann and dilaton equations. Nevertheless, in the 
well-studied case of an exponential potential of the 
form~\cite{lahanas,lahanas2,lahanas3,lahanas4a,lahanas4,plionis}:
\begin{align}\label{dilpotl} 
V\left(\Phi\right)=V_0 \,e^{\tilde \lambda\Phi}, 
\end{align}
with $\tilde \lambda < 0$ a 
constant, 
equation (\ref{dilatoneq}) leads to the lowest-order solution during the 
radiation epoch
 $\Phi \simeq c_{3} \, {\rm ln}a$, implying $\dot{\Phi} \simeq c_{3}H$, where
 $a(t)=a_1 t^{1/2}$, $c_3=-\frac{4}{\tilde \lambda}$ and $a_1=(2\, \tilde 
\lambda^2 V_0)^{1/4}$. Inserting these into 
(\ref{rhodephi})    we find that during the radiation era 
we have
 \be
 \label{1}
\rho_{DE}=M_P^2\left[\frac{6H^2}{\tilde \lambda^2}+\frac{96}{\tilde \lambda} 2^{
\frac { 1 } { \tilde \lambda } } 
 c_{1}   ( \tilde \lambda^2 V_0)^{-\frac{1}{\tilde \lambda}} 
H^{4+\frac{2}{\tilde \lambda}}\right].
  \ee
      Applying it at present time, we can express the parameter $c_1$ in terms 
of 
 $H_0$  and $\Omega_{DE0}$ as
 \be
c_{1}=- \frac{\left(\frac{2}{\tilde \lambda^2}-\Omega_{DE0}\right)\tilde \lambda 
(\tilde \lambda^2)^{\frac
{1}{\tilde \lambda}}H_{0}^{-2}2^{-\frac{1}{\tilde \lambda}}
}{32 } \left( \frac{V_0}{H_0^2}\right)^{\frac{1}{
\tilde \lambda}}. 
\label{c1label}
\ee
Hence, inserting these into (\ref{finalexpress}) we acquire 
\be
\frac{\Delta 
T_{f}}{T_{f}}= \frac{\zeta\left[\frac{2}{\tilde \lambda^2}-\left(\frac{
\zeta
T_{f}^2}{H_{0}}\right)^{\frac{2\left(\tilde \lambda+1\right)}{\tilde 
\lambda}}\left(\frac{2}{\tilde \lambda^2}-\Omega_{
DE0}\right)\right]}{10c_qT_{f}^3},
\ee
with
$\zeta$ given in \eqref{defzeta}.
Interestingly enough we find that the  constraint $\left\lvert\frac{\delta 
T_{f}}{T_{f}}\right\rvert<4.7\times 10^{-4}$ is satisfied  in the narrow window 
\begin{align}\label{lambdawind}
\tilde \lambda \in \left(-1.00561,-1.00558\right),
\end{align}
where we used \eqref{deo}. 
This result was expected, 
since 
in  any model that can describe the  dark energy sector at present times the 
BBN requirements restrict its parameters to suitably narrow ranges, since 
general parameter values would lead to unacceptable large early dark energy 
during the BBN. 

Finally, using (\ref{c1label}) we can find the range of the Gauss-Bonnet 
coupling parameter $c_{1}$,  for given values of the $V_0/H_0^2$, viewing the 
model as 
a phenomenological modified-gravity model,   independent of string theory.
For instance, 
for $V_0/H_0^2\sim10^{94}$ (since $H_0\simeq  
10^{-42}$ GeV  this implies that $\sqrt{V_0}\simeq 10^5$ GeV)
we find $c_{1} H_0^2 \in\left(2.629\times 10^{-95},2.646 \times 
10^{-95}
\right)$, i.e.  $c_{1}   \in\left(1.341\times 10^{-11},1.350 
\times 
10^{-11}
\right)$ GeV$^{-2}$. 

We now remark that, when we concentrate on string theory models~\cite{string,string2,kanti}, we need to use the expression \eqref{c1g} for the Gauss-Bonnet coefficient $c_1$. Taking into account that in standard string phenomenology
$g_s^{(0)2}/4\pi \simeq 1/20$, and that the current collider searches imply 
$M_s > \mathcal O(10)$~TeV, we find  $c_1~\lesssim~10^{-9} $~GeV$^{-2}$. Hence, 
applying the above procedure the other way around, we find from \eqref{c1label} and  
\eqref{lambdawind} that the BBN constraints are satisfied for $V_0/H_0^2 \gtrsim 10^{93}$, which implies 
in order of magnitude the condition $\sqrt{V_0}\gtrsim 10^5$~GeV.
 
In fact, the above results should be interpreted as implying that such dilaton 
dominance in string-inspired cosmologies should end long before the BBN 
era~\cite{spanos}, unless the relevant parameter $\tilde \lambda$ in the dilaton 
quintessence-like potential
lies in the aforementioned narrow window. In that case, however, the dilaton 
presence during BBN might be in conflict with other phenomenological 
consequences of the dilaton cosmology, e.g. supersymmetry searches for dark 
matter at colliders~\cite{lahanas5,spanos,dutta}, provided one attributes the 
dominant dark matter species to supersymmetry. Such issues fall beyond our 
purposes in the current work.

\subsection{$f(P)$ Gravity }

As a first example, namely $f(P)$ Model I,  we consider the power-law model  
 \begin{equation}
 \label{fPmodel}
f\left(P\right)=\alpha P^{n},
\end{equation}
where  $n$ is the only free model parameter, since $\alpha$ 
can be expressed in terms of  $H_0$ and  $\Omega_{DE0}$  given in \eqref{deo} by applying 
(\ref{rhofP}) at the present epoch.
Inserting (\ref{fPmodel}) into (\ref{rhofP}) and then into 
(\ref{finalexpress}) we acquire 
\begin{eqnarray}
&&
\!\!\!\!\!\!\!\!
\frac{\delta 
T_{f}}{T_{f}}=2.1\left(\zeta\right)^{6n-1}\left(T_{f}\right)^{
12n-7}
\nonumber
\\
&&
\cdot\left[\left(-24\right)^{n}-216n\left(n-1\right)\left(-24\right)^{n-1}
+18n\left(-24\right)^{n-1}\right]\nonumber
\\
&&\cdot
\left(30c_q\right)^{-1}\left(6\right)^{1-n}\left(H_{0}\right)^{
2-4n}\left(2H_{0}^2+3\dot{H}_{0}\right)^{2-n}\nonumber
\\
&&\cdot
\left\{\left[
216n\left(n-1\right)-18\right]\dot{H}_{0}H_{0}
^2\right.\nonumber
\\
&&
\left.\ \ \ +54n\left(n-1\right)\left(4\dot{
H}_{0}^2+\ddot{H}_{0}H_{0}\right)-12H_{0}^4\right\}^{-1}.
\end{eqnarray}

\begin{figure}[ht]
\centering
\includegraphics[angle=0,width=0.49\textwidth]{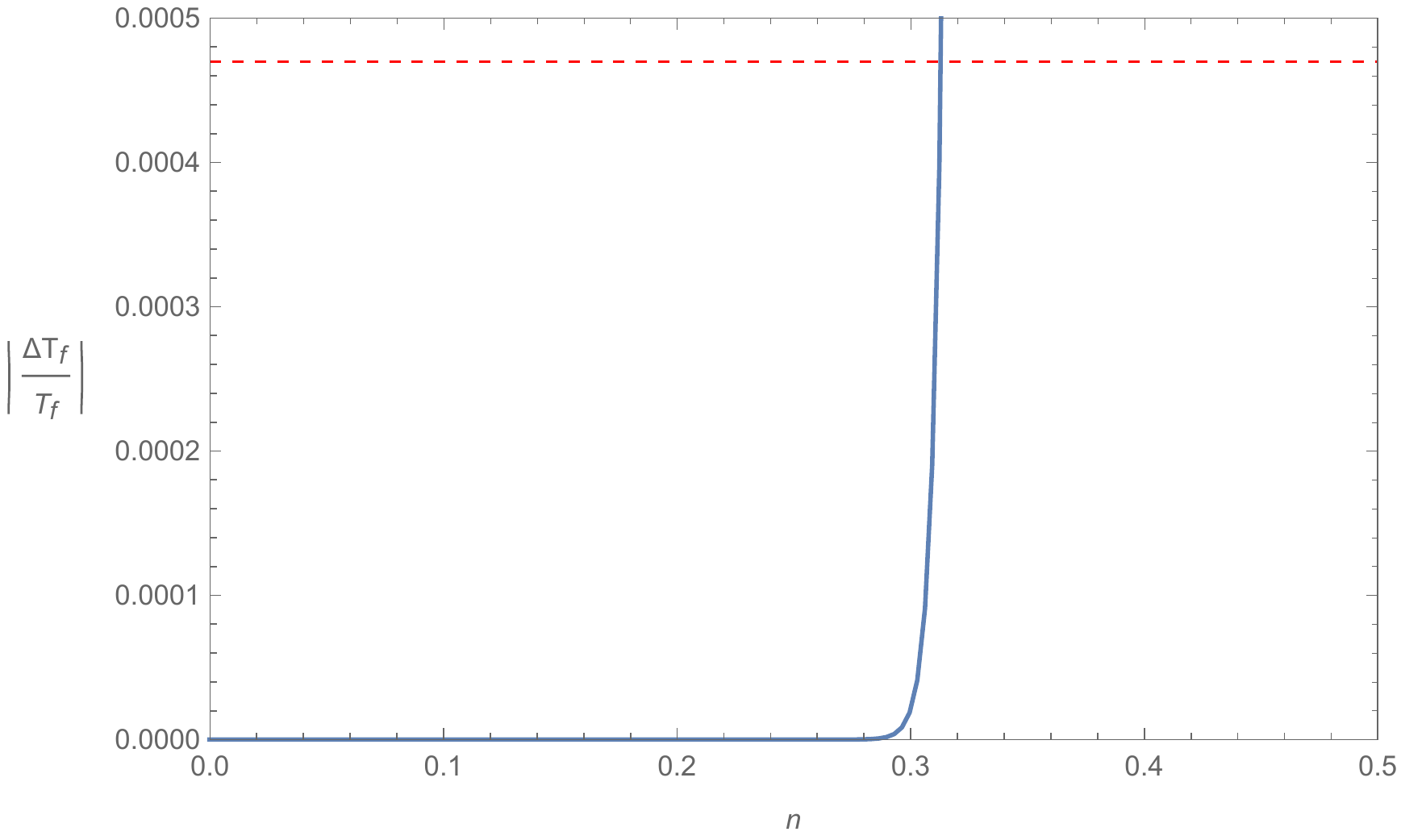}
\vspace{-0.2cm}
\caption{ {
\textit{ $\delta {  T}_f/{  T}_f$ from
(\ref{fG1finalexpr})
vs the model parameter $n$   (blue solid  curve) in the case of $f(P)$ Model   
of (\ref{fPmodel}), and     
the upper bound
for $\delta {  T}_f/{  T}_f$ from (\ref{deltaTbound}) (red dashed line). 
As 
we observe, constraints from BBN require $n\lesssim 0.31$.}}
 }
\label{figfP1}
\end{figure}

In Fig. \ref{figfP1} we draw $\delta {  T}_f/{  T}_f$ from 
(\ref{fG1finalexpr})
 vs the model parameter $n$, as well as the upper bound from 
(\ref{deltaTbound}). It follows from the figure that 
the expression (\ref{fG1finalexpr})   satisfies the bound  (\ref{deltaTbound}) 
of 
$n\lesssim 0.31$.
As we observe,   $n$ is constrained  to small values if we 
want the model to describe dark energy, in which case $\alpha$ and $n$ are not 
independent but related through 
(\ref{rhofP}) at present. If 
we relax this relation    then the BBN 
constraints can always be fulfilled for every $n$ by suitably constraining 
$\alpha$, and   for 
every $\alpha$ by suitably constraining $n$. However, under  their relation, 
i.e.  under the requirement that the $f(P)$ terms 
describe dark energy at the late Universe, then $n$ is constrained close to 
zero, in which case $f(P)$  becomes a constant and the scenario 
becomes $\Lambda$CDM.

\subsection{$f(G)+f(P)$ gravity and cosmology}

For completeness, let us examine a  more realistic case, namely the 
combination of $f(G)$ and $f(P)$ 
gravity. The action is
\be
S=\int d^4x\left[\frac{M_{P}^2}{2}R+f\left(G\right)+f\left(P\right)\right],
\ee
where
\be
f(G)+f(P)=\alpha G^n+bP^n\label{G+P}.
\ee
The dark energy component is
\begin{eqnarray}
&&
\!\!\!\!\!\!\!\!\!\!\!\!\!\!\!\!\!
\rho_{DE}\equiv   
\frac{1}{2}\left[-f\left(G\right)+24H^2\left(H^2+\dot{H}
\right)f^{\prime}\left(G\right)\right.\nonumber
\\
&&\left.
\ \ \ \ \  
-24^2H^4\left(2\dot{H}^2+H\ddot{H}+4H^2\dot{H}
\right)f^{\prime\prime}\left(G\right)\right]\nonumber
\\ \!\!
&& -\left[f(P) +18\tilde{\beta} 
H^4(H\partial_t-H^2-\dot{H})f'(P)\right].
\end{eqnarray}
Using the above analysis for this model we get
\begin{eqnarray}\label{1*}
&&
\!\!\!\!\!\!\!\!\!\!\!\!\!\!\!\!\!\!\!\!\!
\frac{\Delta 
T_{f}}{T_{f}}=-\Omega_{DE0}\zeta^{4n-1}T_{f}^{8n-7}H_{0}^{2-2n}\left(-24\right)^
{n}
\nonumber
\\
&&\cdot \left[ 10c_q\left(a_
{0}-b_{0}\right)\right]^{-1}
\Big\{  
 8n\left(n-1\right) -n    \nonumber
\\ 
 && \ \   +\frac{3}{2}\frac{b}{\alpha}\zeta^{2n}\tilde{\beta}^{n}T_{f}^{4n}
\left[
12n\left(n-1\right) -n \right]\Big\},
\end{eqnarray}
where 
\begin{eqnarray}
&&
\!\!\!\!\!\!\!\!\!\!\!\!\!
a_{0}=\left(24\right)^{n}
\left(n-1\right)\left(H_{0}^2+\dot{H}_{0}\right)^{n-2
}\nonumber \\
&&  
\cdot\left[\left(1-2n\right)\left(\dot{H}_{0}^2+2\dot{H}_{0}H_{0}
^2\right)-n\ddot{H}_{0}H_{0}+H_{0}^4\right],
\end{eqnarray}
and
\begin{eqnarray}
&&\!\!\!\!\!\!\!\!\!\!
b_{0}=108\tilde{\beta}^{n}6^{n-1}H_{0}^{2n}\left(2H_{0}^2+3\dot{
H } _{0}\right)^{n-2} n\left(n-1\right) \frac{b}{\alpha} \nonumber\\
&&
\
\cdot\left[4\dot{H}_{0}H_{0}^2+ 4\dot
{H}_{0}^2+\ddot{H}_{0}H_{0}-12H_{0}^4 \right].
\end{eqnarray}

 \begin{figure}[ht]
\centering
\includegraphics[angle=0,width=0.49\textwidth]{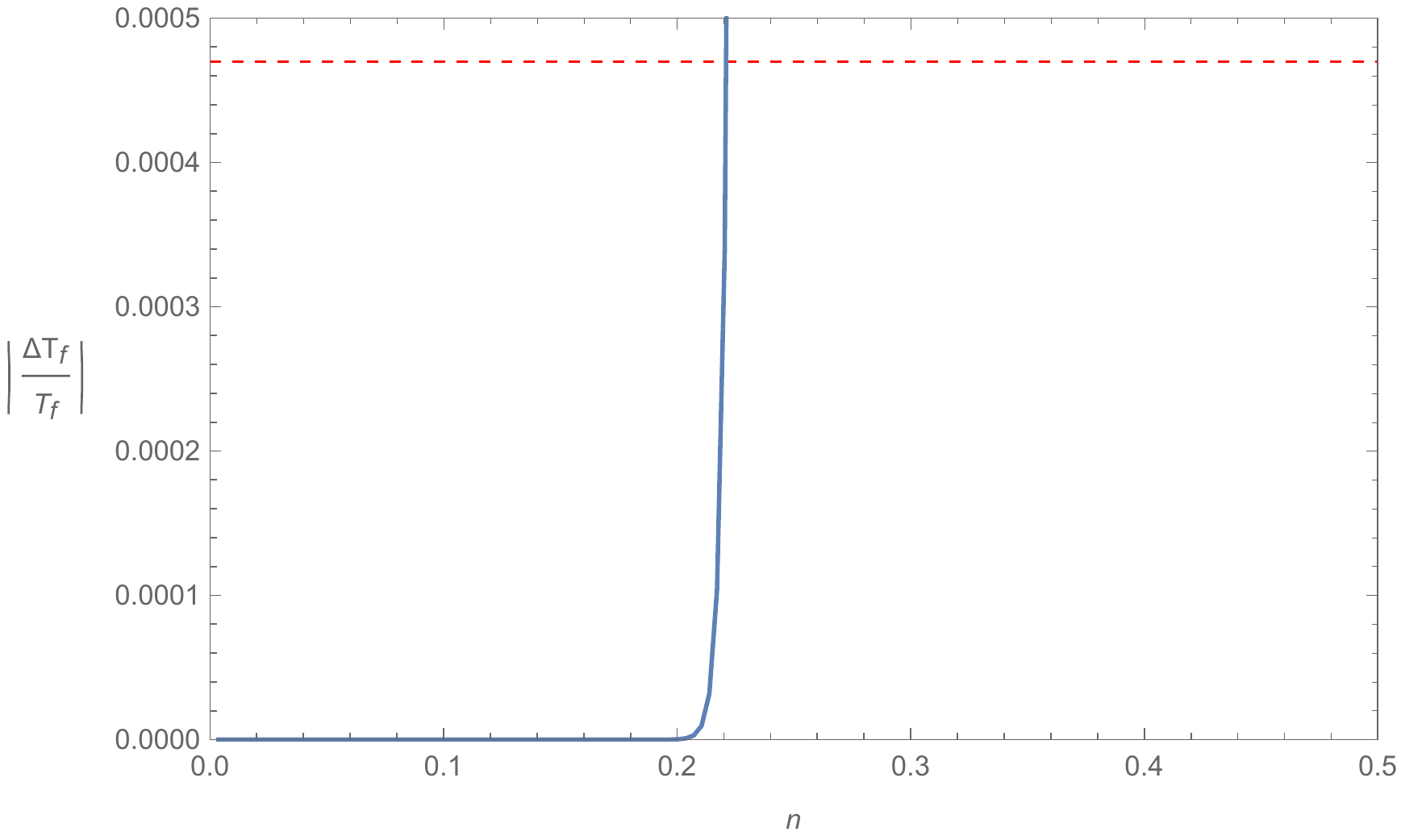}
\vspace{-0.1cm}
\caption{ {
\textit{ $\delta {  T}_f/{  T}_f$ from
(\ref{1*})
vs the model parameter $n$   (blue solid  curve) in the case of $f(G)+f(P)$ 
Model   
of (\ref{G+P}), and     
the upper bound
for $\delta {  T}_f/{  T}_f$ from (\ref{deltaTbound}) (red dashed line). We 
consider the cases $\tilde{\beta}=1$ GeV$^{-2}$  and 
$\frac{b}{\alpha}=10$.
As 
we observe, constraints from BBN require $n\lesssim 0.22$.}}
 }
\label{figfPfG1}
\end{figure}
 
 In Fig.  \ref{figfPfG1}   we draw $\delta {  
T}_f/{  T}_f$ from 
(\ref{1*})  vs the model parameter $n$,   in the case of 
$\tilde{\beta}=1$ GeV$^{-2}$
and 
$\frac{b}{\alpha}=10$,
 as well as the upper bound from 
(\ref{deltaTbound}). We deduce   that  constraints from BBN require $n\lesssim 
0.22$. Similarly, allowing  $\tilde{\beta}$ to vary between   
$ 0.01$~GeV$^{-2}$ and $ 100$~GeV$^{-2}$ leaves the constraint almost unaffected.
Similarly, keeping   $\tilde{\beta}=1$~GeV$^{-2}$ and considering   
$\frac{b}{\alpha}=1000$ leads to $n\lesssim 0.227$, while for 
$\frac{b}{\alpha}=0.001$ we obtain  $n\lesssim 0.21$. In both cases, the 
constraints on $n$ are quite tight.

 \subsection{String-inspired quartic curvature correction models 
\label{sec:qcc}}

 In this subsection we discuss BBN constraints on models containing quartic 
curvature 
 corrections, obtained in the low-energy limit of string theory, \eqref{actionsami}~\cite{sami}.
The first Friedmann equation (\ref{Hrhoc}) can be re-written as
\begin{eqnarray}
\label{Hrhoc*}
6H^2=\rho_m+\rho_r+\rho_{DE},
\end{eqnarray}
in
units where 
$M_P/2=1$,
where we have absorbed all the extra terms in an effective dark energy sector, 
and we have  added the radiation sector for completeness.
 Hence, according to (\ref{corre}), in the case of bosonic strings we have
 \begin{eqnarray}
\label{corre*}
\rho_{DE} &=& B [a_8H^8+a_c I^4+a_4H^4I^2+a_2H^2I^3+a_6H^6I
\nonumber \\
& & -J(a_5H^5+a_1HI^2+a_3H^3I)]\,.  
\end{eqnarray}
 Since in action \eqref{actionsami} the dilaton field does not  evolve 
dynamically, we can choose the same fixed value in the past and present.
Performing the analysis described above    we find
\begin{multline}
\!\!\!\!
\frac{\Delta 
T_{f}}{T_{f}}=\left(60c_q\right)^{-1}\zeta^{7}T_{f}^{9}B\left[a_{8}+a_{c
}+a_{4}-a_{2}-a_{6}\right.\\ \ \
\left. -3\left(a_{5}+a_{1}-a_{3}\right)\right],
\end{multline} 
where $B$ is the free parameter.
For type II strings this satisfies the bound (\ref{deltaTbound}) for
$-2.7\times 10^{-109}\, <\, B \, < \, 
2.7\times 10^{-109}$ or $-3.8\times 10^{-110}\, <\, c_3 
\alpha'^3 e^{-6\phi} \, < \, 
3.8\times 10^{-110} $ in $M_P/2=1$ units. For heterotic type we have 
$-3.6\times 10^{-110}, <\, B \, < \, 
3.6\times 10^{-110} $ or $-6\times 10^{-111}\, <\, c_3 
\alpha'^3 e^{-6\phi} \, < \, 
6\times 10^{-111}$ in $M_P/2=1$ units. Hence, transforming to standard units we 
have $-5.4\times 10^{-37}$  GeV$^{-4}\, <\, B \, < \, 
5.4\times 10^{-37}$  GeV$^{-4}$.

Similarly, for the case of bosonic strings, using (\ref{corre2}) we have
\begin{eqnarray}
\label{corre2b}
\rho_{DE} &=&
A (5H^6+2I^3-6HIJ) +B \{[-21\zeta(3)+210]H^8
\nonumber \\
& &-[3\zeta(3)-90]I^4-[12\zeta(3)+48]H^4I^2
\nonumber \\
&& +
[4\zeta(3)+120]H^2I^3 -[24\zeta(3)-96]H^6I\nonumber \\
&&
+J \{[8\zeta(3)-32]H^5+[12\zeta(3)-360]HI^2
\nonumber \\
&&+
24\zeta(3)H^3I \}\}\,.
\end{eqnarray}
Repeating the above analysis we find
\begin{multline}
\frac{\Delta T_{f}}{T_{f}}=\left(60c_q\right)^{-1}\zeta^5T_f^5\\
\times\left[21A-28B\zeta^2T_f^4\left(\zeta(3)+\frac{285}{7}\right)\right].
\end{multline}
Now, imposing an indicative value for the parameter $B$ from the above range, 
i.e $B= 10^{-109} $ in
$M_P/2=1$ units, 
we find that   the range of $A$ satisfies the bound 
(\ref{deltaTbound}) for $2 \times 10^{-74}\, <\, A \, < \, 
3.1\times 10^{-74} $ or $8.4\times 10^{-76}\, <\, c_2 \alpha'^2 
e^{-4\phi} \, < \, 
1.3\times 10^{-75}$.  Hence, transforming to standard units we 
have $4.1 \times 10^{-38}$ GeV$^{-2}\, <\, A \, < \, 
4.5\times 10^{-36}$ GeV$^{-2}$.
Finally, imposing the indicative value  $B=-10^{-109} $ we find 
$-3.1\times 10^{-73}\, <\, A \, < \, 
-2.0\times 10^{-73} $ or $-1.3\times 10^{-74}\, <\, c_2 \alpha'^2 
e^{-4\phi} \, < \, 
-8.4\times 10^{-75}$, i.e 
$-4.5\times 10^{-36}$ GeV$^{-2}\, <\, A \, < \, 
-2.9\times 10^{-36} $ GeV$^{-2}$.

 
\subsection{Running vacuum  cosmology}

We use the effective dark energy density \eqref{logH2} in the case of (an extended in general) 
running-vacuum type cosmology  in order to evaluate the BBN freeze-out point. We 
start 
by considering the simplest (and more standard) case $ d_{1}=d_{2}=0$. 
It has been argued in \cite{rvmpheno,rvmpheno2,rvmpheno2b,rvmphenot} (see also \cite{tsiapi,tsiapi2}), that at modern eras,
in which the $H^4$ term is not dominant~\cite{rvmevol,rvmevol2}, fitting the corresponding terms of \eqref{logH2}
with the plethora of the cosmological data for the current era, yields 
\begin{align}\label{rangenu0}
0 < \nu = \mathcal O(10^{-3}).
\end{align}
We stress that the presence of the non-zero parameter $\nu$ affects the running 
of the 
matter and energy densities in the current era, leading to observable (in principle) deviations from the $\Lambda$CDM paradigm~\cite{rvmpheno,rvmpheno2,rvmpheno2b,tsiapi,tsiapi2}. 
The cosmological constant term $c_0$ in \eqref{logH2} is then fitted using the relation
\begin{align}\label{c0}
c_0 = H_0^2 \, (\Omega_{DE0} - \nu ),  
\end{align}
where $H_0$ is the current value of the Hubble parameter, and 
\begin{align}\label{de0}
\Omega_{DE0}=\rho_{DE0}/\left(3M_{P}^2H_{0}^2\right), 
\end{align}
with $\rho_{DE0}$ the dark-energy contribution to the 
vacuum energy budget today. This is constrained by the available current-era data~\cite{Planck} to the value 
given in \eqref{deo}.  Note that $\alpha$ does 
not appear in \eqref{c0}, since it multiplies $H^4$ which as we mentioned is 
negligible at late times.

In what follows, we shall concentrate on constraining $\nu$ solely from BBN 
constraints, without further attempts to fit it using the plethora of the 
available data, simply by requiring that the running vacuum energy density is 
associated with the dark energy contribution, which are the basic assumption of 
our analysis. This will provide additional constraints in the range of $\nu$, 
which can then be compared with the value obtained from the analysis of 
~\cite{rvmpheno,rvmpheno2,rvmpheno2b,tsiapi,tsiapi2}.
\begin{itemize}

 \item 
First, we examine models with 
$\frac{\alpha}{H_{I}^2}\equiv c_{2}=d_{1}=d_{2}=0$, $c_{0}\neq 0$, $\nu\neq 0$ 
which are also relevant for the current-era phenomenology, since, as already 
mentioned, the $H^4$ terms in the vacuum energy density \eqref{logH2} are not 
dominant at late epochs~\cite{rvmevol,rvmevol2,rvmpheno,rvmpheno2,rvmpheno2b}. 
Using the constraint \eqref{c0} to eliminate $c_0$, we find
\begin{equation}\ \ \ \
\frac{\Delta T_{f}}{T_{f}}=\left(10\zeta 
c_qT_{f}^{7}\right)^{-1}\!\left[H_{0}^2\left(\Omega_{DE0}
-\nu\right)+\nu\zeta^2T_
{f}
^4\right].
\end{equation}
Hence, on account of \eqref{deo}, the BBN bound (\ref{deltaTbound})   implies that  
\begin{align}\label{nurangeBBN}
-0.0023\lesssim\nu\lesssim 
0.0023~. 
\end{align}
Notice that the satisfaction of the BBN constraints is possible even with negative or zero values of $\nu$. On the other hand, as mentioned above, fitting the plethora of the other modern-era cosmological data requires~\cite{rvmpheno} $\nu > 0$. Nonetheless, the order of $10^{-3}$ for the parameter $\nu$ found here ({\it cf.} \eqref{nurangeBBN}) is in agreement with those fits. Hence, the conventional running vacuum model with a vacuum energy density that fits the current-era constraints~\cite{rvmpheno,rvmpheno2b}, with $\nu =\mathcal O(10^{-3}) >0$, also satisfies the BBN constraints.

 \item 
We continue with the case  $ d_{1}=d_{2}= c_0=0$,  $\nu\neq0$ and 
$\frac{\alpha}{H_{I}^2}\equiv c_{2}\neq0$. Since, as we have mentioned above, we 
do not expect the $H^4$ terms to play any role in the late eras of the Universe 
evolution, we do not expect any strong constraints on  the parameter $\alpha$ in 
this case. 
Indeed, we have
\be\label{2*}
\rho_{DE}=3M_{P}^2H^2\left(\nu +c_{2}H^2 \right).
\ee
 Inserting it in   \eqref{de0}   we 
extract the condition
 \be\label{nuc2}
 \nu =\Omega_{DE0}-c_{2}H_{0}^2.
 \ee
 Using (\ref{nuc2})  to eliminate $\nu$ in terms of $c_{2}$ we finally result to
 \be \ \ \ \ \ \
\frac{\Delta T_{f}}{T_{f}}=\frac{1}{10}\zeta 
T_{f}^{-3} c_q ^{-1}\left[\Omega_{DE0}-c_{2}\left(H_{0}^2-\zeta^2T_{f
}^4\right)\right].
\ee
 Note that this is a linear expression in terms of $c_2$, and thus we easily 
deduce, using \eqref{deo},  that for $c_2\geq0$ (required from the running 
vacuum model)  it does not satisfy the BBN  bound   (\ref{deltaTbound}).
  
 \item 
In  the case  $   d_{1}=d_{2}=\nu=0$,  $c_0\neq0$ and
$\frac{\alpha}{H_{I}^2}\equiv c_{2}\neq0$,  
we have  the constraint
\be
c_{0}=H_{0}^2\left(\Omega_{DE0}-c_{2}H_{0}^2\right),
\label{aux11}
\ee
which can be used to eliminate $c_0$. Hence, concerning $\frac{\Delta 
T_{f}}{T_{f}}$   we finally find 
\be\ \ \ \
\frac{\Delta T_{f}}{T_{f}}=\left(10\zeta 
 c_qT_{f}^{7}\right)^{-1}\left[H_{0}^2\left(\Omega_{DE0}-c_{2}H_{0}^2\right)+c_{
2}
\zeta^4T_{f}^8\right].
\ee
Thus, in this case the bound (\ref{deltaTbound}) leads to 
  \begin{align}\label{c2}
  0\lesssim c_{2}\lesssim 9.7\times 
10^{46}~{\rm GeV}^{-2}~. 
\end{align}
Now inserting the above range of $c_{2}$ into 
(\ref{aux11}), we
find that 
\begin{align}\label{c0}
\frac{c_{0}}{H_0^2}=\Omega_{DE0}-c_{2}H_{0}^2\approx 0.7,
\end{align}
in agreement with the fit of the RVM model to the other cosmological data, performed in \cite{rvmpheno,rvmpheno2}.

 \item 
In the more general case $   d_{1}=d_{2}=0$ with $c_{0}\neq 0$, $\nu\neq 
0$, $c_{2}\neq 0$, we have the constraint
\be
c_{0}=H_{0}^2\Omega_{DE0}-\nu H_{0}^2-c_{2}H_{0}^4,
\label{aix22b}
\ee
which allows us to eliminate $c_0$ in terms of $\nu$ and $c_2$. 
We extract
 \begin{eqnarray}\label{4*}
&&
\!\!\!\!\!\!\!\!\!\!\!\!\!
\frac{\Delta T_{f}}{T_{f}}=\left(10\zeta 
 c_qT_{f}^{7}\right)^{-1}\left[\Omega_{DE0}H_{0}^2+\nu\left(\zeta^2T_{f}^4-H_{0}
^2\right)\right.
\nonumber
\\ 
&&\left.\ \ \ \ \ \ \ \ \ \ \ \ \ \ \ \ \ \ \ \ \ \ 
+c_{2}\left(\zeta^4T_{f}^8-H_{0}^4\right)\right].
\end{eqnarray}
Thus, imposing $-0.0023<\nu<0.0023$ (which was found in a previous case, 
\eqref{nurangeBBN}) we 
conclude that the expression (\ref{4*}) satisfies the bound 
(\ref{deltaTbound}) for  
$0\, <\, c_{2} \, < \, 9.7\times 
10^{46}~{\rm GeV}^{-2}$ ({\it cf.} \eqref{c2}). 
Using the constraint (\ref{aix22b}) we thus find 
$\frac{c_{0}}{H_0^2}=\Omega_{DE0}-\nu-c_{2}H_{0}^2\approx\Omega_{DE0}-\nu$ 
which then gives 
\begin{align}\label{c0new}
0.6977<\frac{c_{0}}{H_0^2}<0.7023,
\end{align}
again in the ball park of the RVM fit to the other cosmological data~\cite{rvmpheno,rvmpheno2}.

\end{itemize}

We proceed with examining the case  where the $\ln(H) H^{2n}$ terms
in \eqref{logH2} are present, namely the cases with $d_1 \ne 0, d_2 \ne 0$.

\begin{itemize}
 \item 
We focus first on $d_1$, i.e.  we consider $d_{2}=\nu=c_{2}=0$, $d_{1}\neq 
0$, $c_{0}\neq 0$. In this case we extract the    constraint
\be
c_{0}=H_{0}^2\left[\Omega_{DE0}-d_{1}\ln\left(M_{P}^{-2}H_{0
} ^2\right)\right],
\label{aux33b}
\ee
which allows us to eliminate $c_0$ in terms of $d_1$.
Hence, we find 
\begin{eqnarray}
&& \ \ 
\frac{\Delta 
T_{f}}{T_{f}}=\left(10c_q\zeta 
T_{f}^7\right)^{-1}\left\{H_{0}^2\left[\Omega_{DE0}-d_{1}\ln\left(M_{P}^{-2}H_{0
} ^2\right)\right]\right.\nonumber\\
&&
\left.\ \ \ \ \ \ \ \ \ \ \ \ \ \ \ \ \ \ \ \ \ \ \ \ \ \ \ \ \ \  +
d_{1}\ln\left(M_{P}^{-2}\zeta^2T_{f}^4\right)\zeta^2T_{f}^4\right\},
\end{eqnarray}
and therefore we deduce that  
\begin{align}\label{d1range}
d_{1}  \in\left(-1.2\times 10^{-5},1.2 \times 
10^{-5}
\right)~.
\end{align}
Inserting the range \eqref{d1range} of $d_{1}$ into (\ref{aux33b})  we find 
\begin{align}\label{c0range}
\frac{c_0}{H_{0}^2}\in\left(0.697,0.703
\right)~,
\end{align}
that is, the presence of non-polynomial $H^2{\rm ln}(H^2 \, M_P^{-2})$ terms in 
the RVM energy density is consistent with BBN for a range of the relevant 
parameter \eqref{d1range}, which affects only marginally the
standard RVM parameters. This is consistent with the fact that such corrections have been argued above to arise from quantum-graviton fluctuations~\cite{ms1,nmtorsion,nmtorsion2}.

 \item 
Focusing on  $d_2$, i.e. considering  $\nu=d_{1}=0$, and $c_{0}\neq 0$, 
$c_{2}\neq0$, $d_{2}\neq 0$, we first find
  the constraint
\be
c_{0}=H_{0}^2\left\{\Omega_{DE0}-c_{2}H_{0}^2\!\left[1\!+\!d_{2} 
\ln\left(M_{P}^{-2}H_{0}^2\right)\right]\right\},
\label{aux44b}
\ee
which allows us to eliminate  $c_0$ in terms of $c_{2}\neq0$ and $d_{2}\neq 0$.
In this case we find
\begin{eqnarray}
&&\!\!\!\!\! \!\!\!\!\!\!\!\!
\frac{\Delta 
T_{f}}{T_{f}}=\left(10c_q\zeta 
T_{f}^7\right)^{-1}
\Big\{
c_{2}\zeta^4T_{f}^8\left[1\!+\!d_{2}\ln\left(M_{P}^{-2}\zeta^2T_{f}
^4\right)\right]
 \nonumber\\
&&
\ \ \ \,    
+
H_{0}^2\left\{\Omega_{DE0}-c_{2}H_{0}^2\!\left[1\!+\!d_{2} 
\ln\left(M_{P}^{-2}H_{0}^2\right)\right]\right\}\!\Big\}.
\end{eqnarray}
Hence, imposing a typical value
 $c_{2}=10^{46}~{\rm GeV}^{-2}$, that was found above ({\it cf.} \eqref{c2}), we deduce that  
\begin{align}\label{d2range}
d_{2}  \in\left(-4.4\times 10^{-2},5.4 \times 
10^{-2}
\right).
\end{align} 
Inserting this range of $d_{2}$ into (\ref{aux44b}) we find again
$\frac{c_0}{H_{0}^2}\approx 0.7$. The reader should notice that, although $H^4$ terms in the vacuum energy density
are not affecting the current-era phenomenology, nonetheless the terms $d_2 H^4 
{\rm ln}( M_P^{-2} H^2 )$ do affect BBN in general, and thus only a narrow 
window \eqref{d2range} (but considerably wider than the corresponding allowed 
range of $d_1$, \eqref{d1range})  is consistent with standard BBN, under the 
assumption that the RVM provides an alternative to dark energy. 

 \item 
Finally, let us examine the cases where both $d_1$ and $d_2$ are not zero and
  $\nu=0$. Thus, we consider 
$d_{1}\neq0$, $d_{2}\neq0$, $c_{0}\neq0$, $c_{2}\neq0$, $\nu=0$. We first find
  the constraint
\begin{eqnarray}
&&
\!\!\!\!\!\!\!\!\!
c_{0}=H_{0}^2\left\{\Omega_{DE0}-d_{1}\ln\left(M_{P}^{-2}H_{0
} ^2\right)\right.
\nonumber\\
&&
\left. 
\ \ \ \ \ \ \ \ 
-
c_{2}H_{0}^2\left[1+d_{2}\ln\left(M_{P}^{-2}H_{0}
^2\right)\right]\right\}.
\label{aux55b}
\end{eqnarray}
Then we find
\begin{eqnarray}
&&\!\!\!\!
\frac{\Delta 
T_{f}}{T_{f}}
=\left(10c_q\zeta 
T_{f}^7\right)^{-1}
\Big\{H_{0}^2\left\{\Omega_{DE0}-d_{1}\ln\left(M_{P}^{-2}H_{0
} ^2\right)\right. \nonumber \\
&&
\ \ \ \ \ \ \ \ \ \ \ \ \ \ \ \ \ \ \ \ \ \ \
\left.-
c_{2}H_{0}^2\left[1+d_{2}\ln\left(M_{P}^{-2}H_{0}
^2\right)\right]\right\}
\nonumber\\
&&  \ \ \ \  \ \ \ \ \
+
\zeta^2T_{f}^4\left\{d_{1}\ln\left(M_{P}^{-2}\zeta^2T_{f}
^4\right)\right. \nonumber \\
&&
\left. \ \ \ \ \ \ \ \ \ \ \ \ \ \ \
+ 
c_{2}\zeta^2T_{f}^4\left[1+d_{2}\ln\left(M_{P}^{-2}\zeta^2T_{f}
^4\right)\right]\right\}\Big\}.
\end{eqnarray}
Hence, taking the typical values  $c_{2}=10^{46}~{\rm GeV}^{-2}$ and $d_{2}=
10^{-5}$ we find  $d_{1}  \in\left(-1.0\times 10^{-5},1.3\times 
10^{-5}
\right)$, and then according to (\ref{aux55b}) we acquire
$\frac{c_0}{H_{0}^2}\in\left(0.697,0.704
\right)$. 
On the other hand, imposing  
$c_{2}=10^{46}~{\rm GeV}^{-2}$ and  $d_{1}\sim 10^{-5}$ we acquire
$d_{2}  \in\left(-8.5\times 10^{-2},1.2 \times 
10^{-2}
\right)$ and then $\frac{c_0}{H_{0}^2}\approx 0.7$.
Hence,  
$d_{1}$ and $d_2$ are restricted in  narrow windows 
in order for BBN constraints to be satisfied. Nonetheless, such windows do not affect significantly the rest of the RVM parameters, which are in the ball-park of the data fits of the standard RVM~\cite{rvmpheno,rvmpheno2}.

\end{itemize}

In summary, (extended) running vacuum scenarios can satisfy the 
BBN constraints, however the corresponding model parameters are constrained 
around their standard,  $\Lambda$CDM  values. This was expected since such 
models are natural extensions of $\Lambda$CDM paradigm, and hence one can always 
 find a parameter region for which the early time behavior is close to 
$\Lambda$CDM evolution.

\section{Conclusions}
\label{Conclusions}

In this work we   investigated the implications of  higher-order modified 
gravity  to the formation of light elements in the early Universe, namely 
on the Big Bang Nucleosynthesis (BBN). Such gravitational modifications are 
proved to be both theoretically motivated as well as phenomenologically very 
efficient in describing the later times evolution of the universe. Nevertheless, 
in order for such scenarios to be able to be considered as viable, one should 
examine that they do not spoil the early universe behaviour, and in particular 
the BBN epoch.

We investigated various classes of higher-order modified gravity and in 
particular models of $f(G)$ gravity,  Gauss-Bonnet-Dilaton gravity,   $f(P)$ 
cubic gravity, and running vacuum cosmology, under the assumption that  these 
models can quantitatively describe the current dark energy sector. In the case 
of the well studied power-law $f(G)$ model, excluding the case where the 
exponent is $n=1$ since it cannot describe dark energy, we found  the 
constraint: $n\le 0.45$. On the other hand, for 
other $f(G)$ models in the literature that include trigonometric functions and 
logarithms, we found that, as long as one imposes the conditions necessary for 
their theoretical consistency, BBN bounds are always fulfilled.

In the case of Gauss-Bonnet-Dilaton gravity with an exponential potential we 
found that the BBN constraints are satisfied in a narrow window for the 
potential exponent, which subsequently leads to narrow constraints on the  
Gauss-Bonnet - dilaton coupling parameter. For the case of $f(P)$ 
gravity with a power-law form, we found  that the exponent is bounded by $n\le 
0.31$.  For the more realistic case of $f(G)+f(P)$ gravity, we found 
that the exponent is bounded in the more narrow window $n\le 
0.22$.  Moreover, for completeness we extracted the constraints on the
string-inspired quartic curvature corrections models, too.  Finally, we 
examined many sub-cases of  running vacuum scenarios and we 
found that they can satisfy the BBN constraints. However, the corresponding 
model parameters are constrained around their standard   $\Lambda$CDM  values,  
which was to be expected, given that these models are extensions of $\Lambda$CDM cosmology.
 
The present analysis shows that models  of  higher-order modified 
gravity, apart from being closer to a renormalizable gravitational theory, they 
can be viable candidates of the description of  Nature too, since they can 
quantitatively account for the dark energy sector and the late-time acceleration 
of the Universe, without altering the successes of the BBN epoch and the 
formation of light elements. Nonetheless, we mention that in most of the 
cases the corresponding model parameters are constrained in  narrow 
windows, which is expected since it is well known that BBN analysis imposes strong 
constraints on possible deviations from standard cosmology. However, even in 
this case the results of the present work reveal the capabilities 
of such constructions and offers a motivation for further investigation, at 
a more detailed level, of the evolution of cosmic perturbations and their role 
in the large-scale structure of the Universe.
   
Finally, we mention that in this work  we did not consider Chern-Simons 
modifications to gravity, involving coupling of axion fields to gravitational 
anomaly  terms~\cite{jackiw,csgravity,bms}. Such terms arise naturally in some 
string theory models, and they can become non trivial in the presence of CP 
violating perturbations around cosmological backgrounds, such as those due to 
gravitational waves \cite{stephon,bms} (although it must be noted that, in the 
string-inspired model of \cite{bms}, such gravitational anomalies are supposed 
to be absent in the post inflationary era, and hence in the epoch of BBN we are 
interested in here). We hope to study such more general cases in future 
works.

\section*{Acknowledgments}
This research is co-financed by Greece and the European Union (European Social 
Fund-ESF) through the Operational Programme “Human Resources Development, 
Education and Lifelong Learning” in the context of the project
“Strengthening Human Resources Research Potential via Doctorate Research” 
(MIS-5000432), implemented by the
State Scholarships Foundation (IKY). The work  of N.E.M is supported in part by the UK Science and Technology Facilities  research Council (STFC) under the research grant ST/T000759/1. 
S.B., N.E.M. and E.N.S. also acknowledge participation in the COST Association Action CA18108 ``{\it Quantum Gravity Phenomenology in the Multimessenger Approach (QG-MM)}''.

\end{document}